# A QUANTIZED STATISTICAL MODEL OF FLOW STRESS AND GENERALIZED HALL–PETCH LAW FOR DEFORMABLE POLYCRYSTALLINE MATERIALS. A TEMPERATURE-DIMENSION EFFECT

*A.A. RESHETNYAK[1]*

*Laboratory of Computer-Aided Design of Materials, Institute of*
*Strength Physics and Materials Science of SB RAS, 634055 Tomsk, Russia*

*Abstract*

A theory of flow stress is proposed, including the yield strength $\sigma_y$ of polycrystalline materials in the case of quasi-static plastic deformations depending on the average size $d$ of a crystallite (grain) in the range of $10^{-8}$–$10^{-2}$ m. The dependence is based on a statistical model of energy spectrum distribution in each crystallite of a single-mode polycrystalline material with respect to quasi-stationary levels under plastic loading, with the highest level equal to the maximal dislocation energy in the framework of a disclination-dislocation deformation mechanism. A statistically calculated distribution of equilibrium scalar dislocation density in each crystallite leads to a flow stress due to Taylor's strain hardening mechanism containing the usual (normal) and anomalous Hall–Petch relations with $\varepsilon=0.002$ for coarse and nanocrystalline grains, respectively, and reaches the maximum at flow stress values for an extreme grain size $d_0$ of the order of $10^{-8}$–$10^{-7}$ m. The maximum undergoes a shift to the region of larger grains for decreasing temperatures and increasing strains $\varepsilon$. Coincidence is established between the theoretical and experimental data on $\sigma_y$ for the materials with BCC (α-Fe), FCC (Cu, Al, Ni) and HCP (α-Ti, Zr) crystal lattices with closely-packed grains at T=300K. The temperature dependence of the strength characteristics is studied. It is shown using the example of Al that yield strength grows with a decrease in temperature for all the grains with $d$ larger than $3d_0$, and then $\sigma_y$ decreases in the nano-crystalline region, thus determining a *temperature-dimension effect*. Stress-strain theoretical curves $\sigma=\sigma(\varepsilon)$ are plotted for the pure crystalline phase of α-Fe with Backofen-Considére fracture criterion validity. The one-phase model of polycrystalline material is extended by including a softening grain boundary phase into a two-phase model, and then by including dispersion (un)hardening. A quasi-particle interpretation of crystallite energy quantization under plastic deformation is suggested. Analytic and graphic forms of the generalized Hall–Petch relations are obtained in the above samples with different values of grain-boundary (second) phase: with small-angle GB and large-angle GB and constant pores. The maximum of yield strength and respective extremal grain size of the samples are shifted by a change of the second phase. The temperature dependence of yield strength in the range of 150-350K (using the example of Al) demonstrates an increase in closely packed nano-crystalline samples for all values $d<d_{\Sigma 1}\sim 3d_{\Sigma 0}$ ($d_{\Sigma 0}<d_0$) with a growth of temperature. An enlargement of the second phase in a sample neutralizes this property (for constant $d$-independent pores). Stress-strain theoretical curves for single-mode and two-mode two-phase PC model of α-Fe are constructed, in comparison with experimental and one-phase model data, and shown to be strongly dependent on the input from multimodality and grain boundaries.

## 1. Introduction

One of the main trends in materials science is a search for controlling the internal defect substructure of crystallites in order to attain the best strength and plastic properties of polycrystalline (PC) materials. An optimization of the above properties is impossible without the benefit of new technologies, the best known of which are the methods of severe plastic deformation (SPD), combined with recrystallization annealing, the vapor deposition method, etc [1]. These technologies allow for ample variations in the orientation and linear size $d$ of the elements of material microstructure, varying from mesopolycrystalline and coarse-grained (CG, 10–1000 μm) to fine-grained (FG, 2–10 μm), ultrafine-grained (UFG, 0.5–2 μm), submicrocrystalline (SMC, 100–500 nm), and further down to nanocrystalline (NC, <100 nm) samples. Experimental research for the physico-mechanical properties of PC materials (microhardness $H$, yield strength $\sigma_y$, ultimate stress $\sigma_S$, and strain hardening coefficient $\theta$) has revealed certain features of the hardening mechanism in the transition to UFG, SMC and NC states for a given material. Systematic research for the influence of the structure parameters of a material on the strength properties under quasi-static deformation was initiated in [2, 3] by the empirical Hall–Petch (HP) relation

$$\sigma_y(d) = \sigma_0 + kd^{-1/2} \qquad (1)$$

($\sigma_0$ and $k$ are the frictional stress in dislocations as they move inside the grains, and the Hall–Petch coefficient, respectively), observed at the initial stage of the yield surface (Fig. 1c) in the diagram for σ=σ(ε) (Fig. 1a) in materials either having grains of different sizes (such as Cu in Fig. 1b), or being at the formal value $\sigma_y(d) = \sigma(d)|_{\varepsilon=0,002} \equiv \sigma_{0,2}(d)$ without any pronounced yield surface. This research was continued in the works of R. Armstrong, H. Conrad, U.F. Kocks, G. Langford, A.W. Thompson, J.G. Sevillano, S.A. Firstov, B.A. Movchan, V.I. Trefilov, Yu.Ya. Podrezov, V.V. Rybin, V.A. Likhachev, R.Z. Valiev, V.E. Panin, E.V. Kozlov, N.A. Koneva, A.D. Korotaev and A.N. Tyumentsev, reviewed in [4, 5]. For UFG, SMC and NC samples, the relation (1) exhibits a significant deviation, which requires a modification [6] of its right-hand side by a term quadratic in $d^{-1/2}$,

$$\sigma_y(d) - \sigma_0 = k_1 d^{-1/2} + k_2 d^{-1}, \qquad (2)$$

and taking into account the parabolicity of the plot ($d^{-1/2}, \sigma_y(d)$), as well as the as the presence of a maximum at the yield strength associated with the so-called "negative value" of the Hall–Petch coefficient $k = (d\sigma_y)/(d(d^{-1/2}))$ in the region of the anomalous Hall–Petch relation. There are quite few models intended to justify the fulfillment of either the (standard) linear or the quadratic Hall–Petch relation, based on empirical approaches. Among them, for example, in [5] the following models were distinguished: the Kocks–Hirth, Arkharov–Westbrook, , Kim–Estrin–Bush, Mughrabi, Ashby, Koneva and Valiev models, the dislocation hardening model, the "casing" model, and the 3D-dimensional composite models. A peculiar feature of these models is grain boundary hardening due to dislocation ensembles, including so-called triple and quadrupole joints of grains, related with their contribution to (1), (2), and also with the concept [8, 9, 10, 11, 12] of increased curvature-torsion in a crystal lattice (CL).

---

[1] e-mail: reshet@ispms.tsc.ru

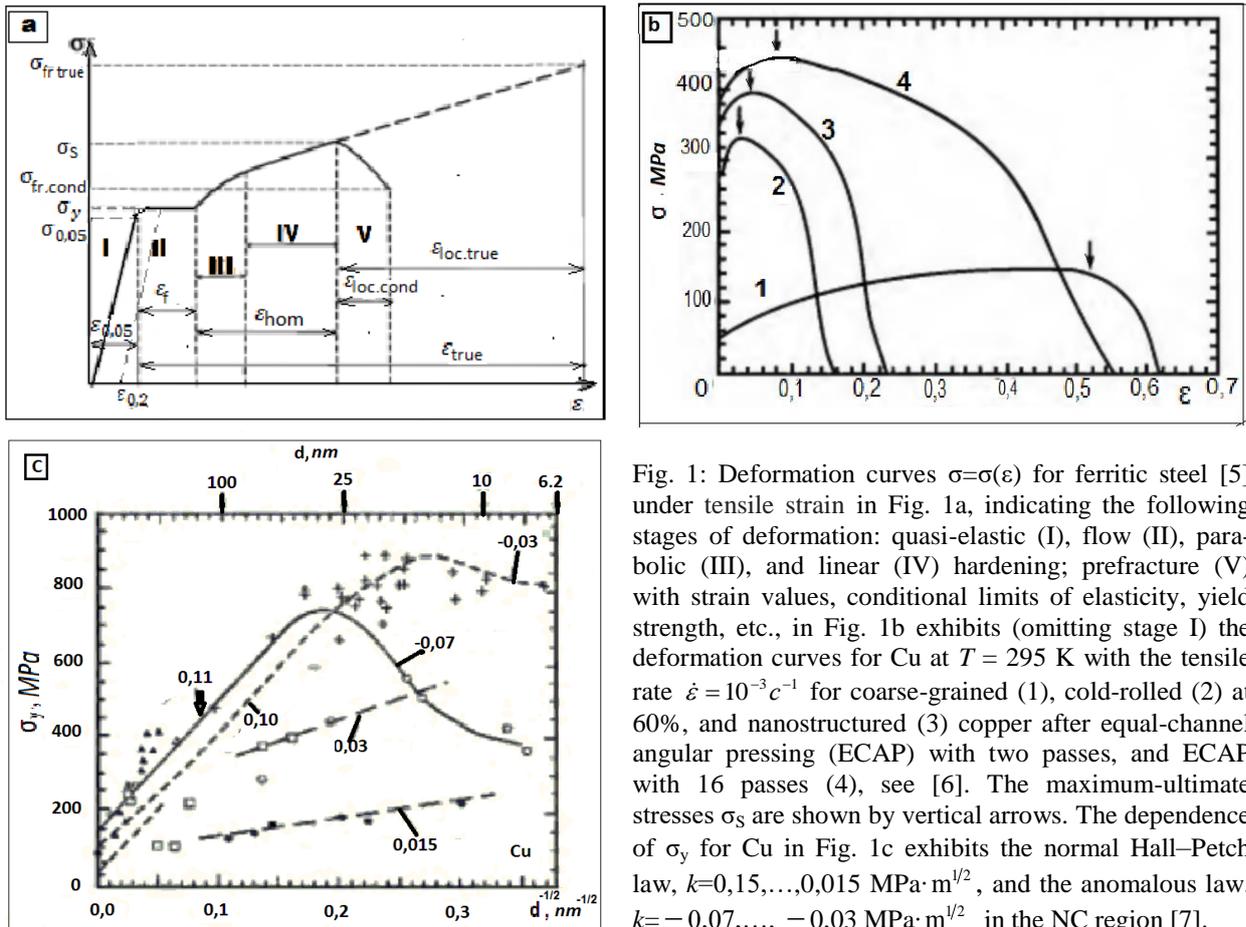

Fig. 1: Deformation curves $\sigma=\sigma(\varepsilon)$ for ferritic steel [5] under tensile strain in Fig. 1a, indicating the following stages of deformation: quasi-elastic (I), flow (II), parabolic (III), and linear (IV) hardening; prefracture (V) with strain values, conditional limits of elasticity, yield strength, etc., in Fig. 1b exhibits (omitting stage I) the deformation curves for Cu at $T = 295$ K with the tensile rate $\dot\varepsilon = 10^{-3} c^{-1}$ for coarse-grained (1), cold-rolled (2) at 60%, and nanostructured (3) copper after equal-channel angular pressing (ECAP) with two passes, and ECAP with 16 passes (4), see [6]. The maximum-ultimate stresses $\sigma_S$ are shown by vertical arrows. The dependence of $\sigma_y$ for Cu in Fig. 1c exhibits the normal Hall–Petch law, $k=0,15,\ldots,0,015$ MPa·m$^{1/2}$, and the anomalous law, $k=-0,07,\ldots,-0,03$ MPa·m$^{1/2}$, in the NC region [7].

In reference to PC aggregates of two-phase materials, the problem of analyzing the behavior of flow stress (FS) as a function of the size of the grain (which is the main solid phase) and as an effect of grain boundaries at the soft (second) phase, becomes more involved (with the contribution of the soft phase increasing to tens of percent in the transition to SMC and NC materials [13]) and was examined for metal, metal-ceramic and ceramic materials in the review [4]. Since the production of a uniformly sized (single-mode) grains of materials is technologically difficult, this leads to making allowance for distributions with respect to the grain size in a sample, and thereby also takes into account the specifics of calculations for plastic and strength parameters, in particular, FS and $\sigma_y$. For such samples, beyond the relation (2) for FS and $\sigma_y$ in the case of lognormal grain size distribution, a different model of dependence on the grain size was proposed [14, 15] for samples coated with Cr by magnetron sputtering [16]. The model takes into account a deviation from the strictly quadratic dependence (2) for $d<d_{cr2}$, by using the S-integrals technique combining three relations: equation (1) for $d_{cr1}<d$ ($d_{cr1}=(k_1/k_2)^2 \leq 0.5$ μm, equation (2) for $d_{cr2} \leq d \leq d_{cr1}$ ($d_{cr2} \leq 0.1$ μm), and the new relation (for $d<d_{cr2}$):

$$\sigma_3 = \left(1-\left(\frac{d-t}{d}\right)^2\right)\sigma_{gb} + \left(\frac{d-t}{d}\right)^2 \sigma_{rh}, \qquad (3)$$

where $t$, $\sigma_{gb}$, $\sigma_{rh}$ are the thickness, ultimate stress of the grain boundary, and theoretical ultimate stress of a grain, respectively. Since we assume a bi-quadratic dependence on $d^{-1/2}$ for large values such as ($\sigma_{gb}$; $\sigma_{rh}$) = (2;12) GPa, the model allows us to go over to the NC region, where the anomalous (inverse) Hall–Petch law holds true [17–21], with a decrease in FS and $\sigma_y$ as $d^{-1/2}$ increases for $d<100$ nm.

Among the theoretical models leading to a simultaneous description of the normal and anomalous Hall–Petch laws for $\sigma_y$, as well as microhardness $H$, attention is due, first of all, to *a mixed model of plasticity in polycrystalline metals, supplementing dislocation plasticity inside the grains by a mechanism of slipping along the grain boundaries*, based on the Maxwell strong viscous liquid within a molecular dynamics simulation for Cu and Al [22], and, secondly, to the dislocation kinetic model of G.A. Malygin [23, 24] based on a first-order evolution equation for the average dislocation density $\rho=\rho(t)$ in a grain,

$$\frac{d\rho}{d\gamma} = \frac{\beta}{bd} - (k_a + k_b)\rho, \quad k_b = 4\eta_b \frac{D_{gb}}{\dot{\gamma}d^2}, \quad \eta_b \approx \frac{Gb^3}{k_B T}, \qquad (4)$$

following the Taylor strain hardening mechanism [25]. In obtaining (4), one assumes [23, 24] that the time dependence $\rho=\rho(\gamma(t))$ is implicit through the uniaxial tensile strain (or compression) $\varepsilon=\gamma/m$, $d\rho/dt = \dot{\gamma} d\rho/d\gamma$, for a constant strain rate $\dot{\varepsilon} = \dot{\gamma}/m$ and shear strain rate $\dot{\gamma} = b\rho u$, where $b$ is the module of the Burgers vector, $u$ is dislocation velocity, $m=3.05$ is the Taylor orientation factor, and ($\beta, k_a, k_b; D_{gb}, G, k_B, T$) are, respectively, the coefficients determining the intensity of dislocations accumulated in a grain volume and the annihilation of screw and edge dislocations, the grain-boundary diffusion coefficient, the shear modulus, the Boltzmann constant, and absolute temperature. The model implements a competitive process of proliferation and annihilation of dislocations which depends on a sufficiently large number of external parameters. Finally, one also considers some models with 3D dynamics of discrete dislocations [26, 27, 28].

The general conclusions drawn from the theoretical and experimental research known to date with respect to FS and $\sigma_y$ are as follows:

1) the maximum of $\sigma_y$ is attained in some materials at certain values of the crystallite (grain) diameter $d_0$ in the NC region at a given $T$ and plastic deformation (PD) rate $\dot{\varepsilon}$;
2) $d_0$ is shifted to the region of larger grains with increasing values of $T$ and, independently, with decreasing values of $\dot{\varepsilon}$;
3) in the regions of coarse and NC grains, there is no physical model describing simultaneously the normal and anomalous HP laws, based on a statistical approach to the spectrum of crystallite energies considered as the main (solid) phase of PC materials with a fixed PD, depending on the distribution of dislocation ensembles.

The ongoing discussion concerning the options for producing 1D defects (dislocations) as emerging from 0D defects (in particular, nanopores, vacancies and other zones of localized deformation), due to the lack of unambiguous interpretation of experimental data makes it possible to state that there is no rigorous well-grounded theory taking into account the defect substructure of a CL that would lead to a Hall–Petch-type relation for all the grain ranges in a PC material subject to PDs. Twin type defects (2D- defects), prevailing in NC materials are always produced by those dislocations that can be presented as combinations of dislocations. It should be noted that the cases of the normal (CG materials) and anomalous (SMC and NC materials) Hall–Petch relations actually correspond to the radiation of an absolutely black body, exhibiting the Rayleigh–Jeans (long-wave) and Wien (short-wave) regions in the plot ($\omega,u(\omega,T)$) for the spectral density of radiation energy $u(\omega,T)$ (with the dimensionality $[u(\omega,T)]=[\sigma_y]\cdot 1s=1eV\cdot 1s\cdot m^{-3}$), unit-ed in the framework of Planck's theory [29] based on the discreteness of radiation energy spectrum for atoms in an absolutely black body.

The aim of this work is to construct a theoretical model for the emergence and evolution of a defect structure, including 0D (nanopores, bi-nanopores, etc) and 1D (dislocations) defects in the grains of a loaded PC aggregate, based on a statistical approach to the energy spectrum of a grain, in view of the integral nature of FS and $\sigma_y$. The latter point is of crucial importance due to the overwhelming complexity of a direct solution of the Schrödinger equation (in partial derivatives of order no less than $3 \cdot 10^{12}$) for a crystallite in an external mechanical deformation field of $N=10^{12}$ atoms (corresponding to $d \sim a \cdot N^{1/3} = 3$ μm at the lattice constant $a=0.3$ nm), even with the benefit of advanced supercomputers.

In the present analysis, we consider (Sec. 2) a scenario for the emergence from a sequence of 0D defects (nanopores, localized deformation zones) of an edge dislocation, with estimation of the energies of different dislocations. In Section 3, a model is introduced for the distribution of crystallite energy in a polycrystalline single-mode material with respect to quasi-discrete levels in a state of thermodynamic quasi-equilibrium at a fixed value ε of PD. In Section 4, we obtain the relation for equilibrium (e.g. for annealed materials) FS, as well as the generalized HP law for equilibrium $\sigma_y$, and also study the corresponding asymptotes for CG and NC PC samples.

In Section 5, we introduce a quasi-particle interpretation for the quanta of PD energy set equal to the energy $E_d^{L_e}$ of a unit dislocation necessary for dislocations (including nanopores) to emerge within the scenario for the emergence of edge dislocations starting from a sequence of 0D-defects. The validity of the generalized HP law for the yield strength of a number of single-mode PC materials with different CL in the crystallite phase is verified using graphic representations (Sec. 6). In Section 7, we study the temperature dependence of $\sigma_y$, exhibiting the *temperature-dimension effect* in the example of single-mode PC Al.[2]. In Section 8, we construct stress-strain curves $\sigma = \sigma(\varepsilon)$ for different grain sizes in the first (crystallite) phase of single-mode PC aggregate in α-Fe and study the hardening and pre-fracture stages. The inclusion of a second grain boundary phase leads to the construction of a realistic *composite-like* model that allows one to control the defect structure of both phases in Section 9. In Sections 10, 11, 12, respectively, we employ a two-phase model to study, first, the HP relations for a number of PC materials of Sec. 6 (albeit having different second-phase values), second, together with an extension of the results of Sec. 7, the temperature dependence in the range of 150– 350 K for a two-phase Al, third, the construction of stress-strain curves for single-mode and two-mode PC samples α-Fe with different grain boundaries**.** The work is devoted to the solution of the above problems and is followed by some conclusions.

A grain is understood as a crystallite with an initial (prior to PD) density of dislocations.

## 2. Emergence scenario for deformation-induced dislocations and properties of dislocation energy

In order to formulate our model, we introduce a definition that permits a uniform description of 0D (zero-dimensional) and 1D (one-dimensional) CL defects, using a representation reminiscent of the Frenkel–Kontorova model, proposed in the 1930s, with 0D-defects "rarefaction dislocation" type ("holes" in a CL). For models, we choose non-metallic and metallic solids with a cubic CL. In the former case, one of the six (e.g., covalent) bonds between the atoms connecting an atom in a CL node with the other atoms is caused by an elementary PD act due to tensile strain to undergo a rupture (breaking a common electron pair in the outer electron shell of two other atoms), thus significantly displacing the two atoms participating in the deformation with the emergence of a 0D-defect, a *nanopore*. A nanopore (as a kind of stacking fault of CL) represents a localized deformation (or plasticity) zone in the given 0D-dimensional case. Such an emergence seems to be natural within the thermo-fluctuation mechanism, under which the collective oscillations of atoms (beyond the elasticity limit of a sample) are such that one of the antinodes (as they interfere) accumulates an amount of energy which is larger than the bonding energy between the atoms and is sufficient to create a local stress value being higher than the Peierls–Nabarro stress in the crystallographic plane containing these atoms. Specific locations of nanopore emergence are random. The most probable event is the emergence of a nanopore at the surface of a sample (from the grain boundary). Such an antinode may be born in the region where two or more atoms are localized, thus leading to a disconnection of two and more bonds, and thereby to the emergence of a large nanopore (n-nanopore,

---
[2] E.g. for Pb, the dependence of $k(\varepsilon) = f(T)$ was studied in [30]; for another materials, see the review [5]

n=2,…). For a single-layer material, dislocations can occur only in the layer plane (e.g., for graphene with a hexagonal CL they can be generated by a pair of Stone–Wales defects when breaking the zigzag symmetry [31]), whereas for dislocations (perpendicular to the layer) this is the degenerate case of a pair of dislocations, each containing an atom in its axis (0D-defect) and having no specific Burgers vector (see Fig. 2a for a cubic CL). For a two-layer material (e.g., AB- or AA-stacked bilayer graphene), the mechanical power supplied under a PD to a unit cell and sufficient to break the bonds in the layer between two atoms, as well as the bonds in the other layer between the two atoms combined by orthogonal projection (Fig. 2a), produces the emergence of two unit edge dislocations with parallel axes of length $L=a$, which is the lattice constant being the modulus $b$ ($b=a$) of the Burgers vector for each of the dislocations.

An example scenario for the emergence of a pair of edge dislocations in a crystallite subject to an elementary plastic deformation act consisting of a sequence of single plastic deformation sub-acts under the thermal-fluctuation mechanism with intermediate nanopores in a crystallite sample of a cubic CL is given by Fig. 2.

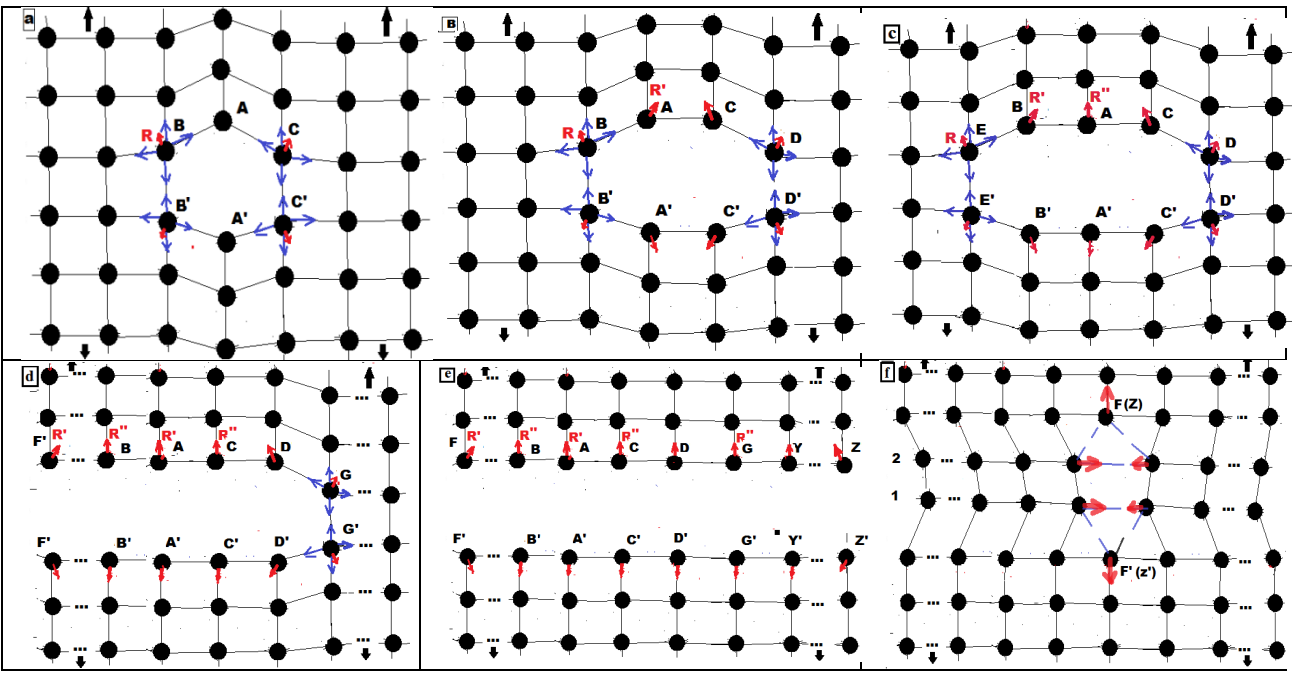

Fig. 2: The emergence process for a pair of rectilinear edge dislocations starting from a growing nanopore (under development) upon stretching along $AA'$ in the crystallographic slip plane. In Fig. 2a, for a nanopore formed at the discontinuty between the nodes $A$ and $A'$ it is shown that in the nodes $B$, $C$ ($B'$,$C'$) neighboring with $A(A')$ there is a formal (from quasi-classical viewpoint) resultant Newton force $R$ indicated in red (additional to the external force and equivalent to tensile stress), leading to the emergence of a bi-nanopore in Fig. 2b. For the same reasons, the 3-nanopore in Fig. 2c is formed and then $n$-nanopore – with the exit of the boundary points $F$ and $F'$ of an $n$-nanopore to the surface from the left in Fig. 2d and then to the surface to the right of the boundary points $Z$ and $Z'$ – there is an $m$- nanopore ($m>n$) in Fig. 2e, with the emergence of a pair of edge dislocations having the length $L=mb$ and opposite Burgers vectors. Also, in Fig. 2f, in the direction perpendicular to the plane of Figs. 2a–2e, the atomic half-planes bounded from inside by the dislocation axes move apart under tension, i.e., the dislocations propagate with their edges exiting the crystallite surface, while the adjacent atomic planes parallel to these half-planes are drawn into the empty space due to inter-atomic forces, thus forming the defect packaging (subtraction). Depending on the magnitude of thermal fluctuations, a $k$-nanopore, $k>1$, can initially be generated at any location in a grain, most probably occurring at the surface, as in the case of a CG sample, see Eq. (15) and the comments below

For metallic PC samples, dislocations can occur in atomic planes preferably along the directions having the least energy of defect packaging (subtraction or insertion), according to a scenario similar to the one presented in Fig. 2 (including BCC, FCC, HCP CL specifics), with a preliminary emergence of nanopores, as a zone of localized plasticity, also known as a band of localized deformation.[3]. In the case

---
[3]In a zone of localized plasticity under a high-level local internal stress, so-called martensitic transformations can occur, providing direct transitions in austenitic steels from the FCC γ-phase of CL at the BCC α(α′)-phase, and

of a pronounced dislocation structure, new dislocations inside a grain may terminate at already existing dislocations in the grain, whereas in the flow region of PC materials the pole Frank–Read mechanism [33] of generating the dislocations, especially in CG—UFG materials, is valid, having a zone of localized plasticity as a necessary condition for driven dislocations to proliferate, starting from a given one.

The time for the nanopore in Fig. 2a to emerge at the PD rate $\dot{\varepsilon}$ is estimated as $t_0 = b/(\dot{\varepsilon} l)$, where $l$ is the length of the crystallographic plane along $AA'$, and $b = a$ is the interatomic distance. For $(\dot{\varepsilon}; b; l) = (10^{-3}\,\text{s}^{-1}; 0,3\,\text{nm}; 10^{-3}\,\text{m})$, the estimate $t_0 = 3 \cdot 10^{-4}\,\text{s}$ holds true. Further, after a short interval, $\Delta t = b/v_s \ll t_0, \Delta t \sim 10^{-12}$ sec, of relaxation to a new equilibrium position (with $v_s$ being the speed of sound in the crystallite), the emergence of a bi-nanopore (Fig. 2b) is more advantageous than the emergence of a nanopore in a different location of the plane, since the atoms adjacent to $A(A')B$, $C(B',C')$ experience some resultant forces additional to the external ones (equivalently, the gradient of tensile stress). Thus, the process of *n*-nanopore emergence from the initial nanopore spreads rapidly up to the boundary points of the $F$, $F'$ plane (Fig. 2d) and is then followed (Fig. 2e) by the emergence of a pair of edge dislocations in the axes $FZ$, $F'Z'$, whose Burgers vectors are opposite and perpendicular to the plane of the figure. Then, the dislocations diverge (Fig. 2f), followed by a characteristic collapse of the neighboring planes parallel to the plane of Figs. 2a–2e, due to a permanently tunable spectrum of energy levels (as dislocations propagate) for the atoms in these and neighboring atomic planes, thereby admitting some new stable (e.g., according to the Landau–Zener mechanism) interatomic bonds with a new electronic structure. As a result, when the forces that bind the atoms of the dislocation axes with the atoms of the neighboring planes become stronger than the stretching PD forces, the usual picture is reproduced for a unit dislocation [34], with a far-gone stationary second dislocation. After the emergence of an *m*-nanopore, the shear perpendicular to the PD direction may enter into competition, instead of tension, with the respective change of grain orientation, for which the dislocation axes $FZ$ and $F'Z'$ are shifted relative to the direction of tension at the different sides of Fig. 2e.

Some remarks are in order. First of all, the instantaneous emergence of a large dislocation under a PD without any intermediate 0D-defect in a multilayer crystallite is in contradiction with the finiteness of the interaction velocity. Second, we select two kinds of time scaling: the fast scaling $t_d \approx N\Delta t$ for a dislocation emergence starting from a sequence of nanopores, and the slow scaling proportional to $\hat{t}_0 = b/(\dot{\varepsilon} d)(t_d \ll \hat{t}_0 < t_0)$ for enumerating the PD acts. Thus, from the slow-time scaling we may say on instantaneous emergence of a dislocation within the scenario above. Third, an experimental confirmation of dislo-cation emergence on the basis of a given sequence of nanopores requires some precise measurements in view of the transience of dislocation process, and also due to the blurring (justifying the emergence of na-nopores) of a diffraction pattern due to the screening of the plane containing the nanopores by the neighboring parallel atomic planes. Fourth, the above scenario for the emergence of a screw dislocation followed by a mixed dislocation may also be investigated (we leave this problem outside the paper's scope).

The above analysis makes it natural to extend the concept of dislocations (introduced by V. Volterra in 1905, followed by E. Orowan, M. Polanyi and G. Taylor in 1934 for edge dislocations, and afterwards by J. Burgers in 1938 for screw dislocations) by a definition due to F.Ch. Franck (see, for instance, [34, 35, 36]), according to which a subsequent (second) dislocation in the same crystallographic slip plane with the opposite Burgers vector is far removed by the action of loading (stretching).

We refer to a *generalized dislocation* (GD) with its axis (of length $na$) consisting of ($n+1$) atoms ($n$ segments) as a topological defect of physical spatial dimension, $D$, $D \leq 1$, for which there exists at least one closed Burgers contour around its axis at the distance of no less than the *a*-atomic lattice constant, which determines a Burgers vector **b**, being constant along the axis (line) of a generalized dislocation, with a possible exception of the end points.

The rule for determining the direction and magnitude of the Burgers vector remains the usual one, according to the "right screw" [34]. When the GD ends exit to the boundary of a crystallite, or when they coincide (the emergence of a loop), we have a usual dislocation (edge, screw, or mixed type). Otherwise, the GD represents an incomplete dislocation of one of these types if there exist more than one crystal lattice nodes in its axis, with a Burgers vector in the dislocation axis which is undetermined only at its finite

---

then inverse transitions, with forming the 2D twin type defects or dislocations by means of combinations of partial Shockley dislocations, , taking into account a change of the shear direction under direct and inverse transitions, justified empirically in [32].

points, or else, if the dislocation, whose ends are identical, represents a 0D-defect being a nanopore in the limiting case of dislocation. An incomplete dislocation in whose axis there are *n* atoms (*n*>2) always has a closely-situated second incomplete dislocation, thus resulting in an *n-nanopore* (see Figs. 2a–2d for *nanopores*, *bi-nanopores*, *3-nanopores* and *n-nanopores*). A rectilinear *n-nanopore* implies the presence of *n* "holes", being empty CL nodes in the interval. Such a sequence of 0D- defects actually contains two axes of *n* atoms each (*ED*, *E′D′* in Fig. 2c; *FG*, *F′G′* in Fig. 2d), being parallel and spaced by the distance 2*a*, except the ends (*E* and *E′*, *D* and *D′*, *G* and *G′*), spaced by the distance *a*. It is such axes of incomplete dislocations that we understand as the axes of two GDs characterizing an *n*-nanopore from a zone of localized deformation.

Dislocations create elastic stress fields with a tensor $\sigma_{ik}, i,k = 1,2,3$, which define the field of elastic strain with a tensor $u_{ik}$ in a crystallite with a shear modulus *G*, so that the analytically free energies of screw, edge, and mixed dislocations of length *L* with the Burgers vector **b** in the crystallite are calculated by the rule [34] (with the free energy *F*), $F = 1/2 \int \sum_{i,k=1}^{3} u_{ik}\sigma_{ik} dV$, respectively,

$$E_d^{screw} = \frac{Gb^2 L}{4\pi}\ln\left(\frac{R}{r_0}+Z\right), \quad E_d^{edge} = \frac{Gb^2 L}{4\pi(1-\mu)}\ln\left(\frac{R}{r_0}+Z\right), \quad E_d^{mix} = \frac{Gb^2 L}{4\pi K}\ln\left(\frac{R}{r_0}+Z\right). \quad (5)$$

Here, $\mu$, *R*, $r_0$, *Z* are, respectively, the Poisson ratio of the material (0.1 <μ <0.4), the radii of the dislocation zones (the cut-off parameter *R*, usually, $R=n\cdot 10^4 b$) and of the dislocation core (axis) $r_0 \approx 3b$, the correction constant $1 < Z < 3$ for estimating the energy near the dislocation core, $(1-\mu) \leq K \leq 1$.

The energies of edge and screw dislocations are nearly the same and have the following properties:
1. the energy of a dislocation is proportional to the dislocation length: $E_d^{mix} \sim L$;
2. the energy of a *unit dislocation* of length $L_e = b$ at $R \sim L_e$ equals to $E_d^{L_e} = \frac{1}{2}Gb^2 L_e = \frac{1}{2}Gb^3$; and for a PC material with (*G*,*b*)=(30 GPa, 3·10$^{-10}$ m) it equals to $E_d^{L_e}(G,b) = 2{,}53\,\text{eV}$;
3. the dislocation energy containing (*n*+1) atoms (for complete dislocation) in the dislocation axis[4] equals to the sum of the energies of *n* unit dislocations $E_d^L = nE_d^{L_e}$;
4. a dislocation with a smaller module of the Burgers vector $b_1$ of the same length *L* as a dislocation with $b = mb_1$ is more advantageous energetically at its emergence, since $E_d^L(G,b) = m^2 E_d^{L_e}(G,b_1)$;
5. **the energy of a unit dislocation exceeds by two orders the energy of thermal fluctuations of an atom, $k_B T$ at T = 300K: $E_d^{L_e}(G,b)/k_B T = (2{,}53\,\text{eV})/(0{,}026\,\text{eV})$;**
6. the energy $E_d^{L_e}$ with the smallest Burgers vector is comparable with the activation energy of an atom $E_d^{L_e} \approx E^{act}$ in the course of diffusion. Indeed, the diffusion coefficient reads $D = D_0 e^{-(E^{act}/k_B T)}$, with the frequency factor $D_0$ for most of the metals[37]. being $E^{act} \in [1\,eV, 4eV]$, so for copper, α-iron, niobium, we have $[E_d^{L_e}, E^{act}](Cu)$=[2.31;2.05] eV; $[E_d^{L_e}, E^{act}](Fe)$=[3.93;3.05]eV; $[E_d^{L_e}, E^{act}](Nb)$ =[4.21; 4.13] eV at *T*=300*K* and (*G*;*b*)=(37.5 GPa; 3.30·10$^{-10}$m), for Nb with BCC lattice at *b* equal to its CL constant (for Cu and α-Fe, see Table 2 [39]);
7. for a crystallite being a polyhedron of diameter *d* inscribed in a sphere, the largest rectilinear dislocation lies in one of the equatorial slip planes passing through the center of the crystallite, and the largest loop dislocation coincides with the equator of the polyhedron slip plane (Fig. 3), having the respective length and energy

$$(L; L_l) = (1; \pi)d = (N; \pi N)b, \quad E_d^N = \frac{Gb^2 d}{4\pi K}\ln\left(\frac{R}{r_0}+Z\right)(1;\pi) = \frac{Gb^3 N}{4\pi K}\ln\left(\frac{R}{r_0}+Z\right)(1;\pi), \quad (6)$$

where $N = [d/b; \pi d/b]$ is the number of atoms on the corresponding dislocation axes, and the square brackets denote the integer part of the ratios *d/b* and π*d/b*.

---
[4] The axes of partial dislocations do not need to contain atoms, but they do consist of elementary segments whose lengths are proportional to their Burgers vectors. If stated otherwise, we discuss complete dislocations alone.

In the first place, it follows that an *n*-nanopore and two parallel dislocations of *n* atoms in their axes are comparable energetically. Secondly, it is advantageous to realize dislocation ensembles under PD by corresponding crystallographic slip systems with the smallest Burgers vector ***b***, including partial dislocations, especially for materials with FCC or BCC lattices. On the basis of property 6, one can make an approximate assumption to the effect that the energy of an arbitrary dislocation may be estimated analytically using the activation energy (determined experimentally) of the atoms that form the axis.

### 3. Statistical model of crystallite energy distribution under quasi-static PDs

Consider a polycrystalline single-mode metal aggregate of volume *V* with an arbitrary CL, being homogeneous with respect to the size of crystallites closely packed in the form of polyhedra (of diameter *d*) distributed isotropically throughout the sample. The PC aggregate is taken in a fixed phase state being constant within a considerable range of temperatures [$T_1$,$T_2$], see Fig. 3. We restrict ourselves to the case of a cubic CL with the smallest Burgers vector of an arbitrary dislocation coinciding with *a*, $b=a$.

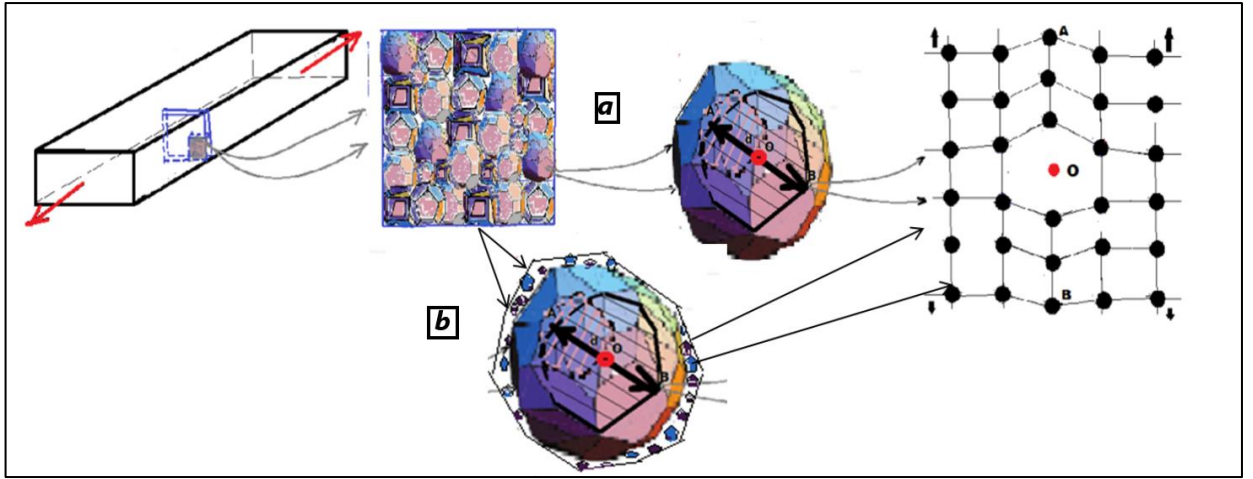

Fig. 3: Multilevel models: a one-phase polycrystalline sample with uniformly sized (single-mode) crystallites, labeled by ***a***; a two-phase polycrystalline sample with uniformly sized (single-mode) crystallites with crystallites and porous from the second phase (grain boundary region) around the basic crystallite, labeled by ***b***, and a crystallographic slip plane passing through the crystallite center.

Let the process of quasi-static mechanical loading (stretching) of a sample with a constant strain rate, $\dot{\varepsilon}, \dot{\varepsilon} \in [10^{-5},10^{-3}] \text{s}^{-1}$, begins at the time instant $t_0=0$ and is characterized by the temperature *T*. When the elastic limit $\sigma_e$ is reached, with the conditional value $\varepsilon_{0.05} = 0,25 \cdot \varepsilon_{0.2}$ of limiting elastic deformation, a PD starts to emerge in crystallite nanopores and dislocations, accompanied by energy exchange between the atoms released from the CL nodes and the nanopores caused by the CL breaking. When the residual PD $\varepsilon_{0.2} \geq \varepsilon > \varepsilon_{0.05}$ (corresponding for $\varepsilon = \varepsilon_{0.2}$ to $\sigma_y$) is reached at the instant $t = \varepsilon/\dot{\varepsilon}$, $t > t_1 = \varepsilon_{0.05}/\dot{\varepsilon}$, with a fixed external loading ($\dot{\varepsilon}=0$) the state of thermodynamic quasi-equilibrium (see footnote 5) is established for each crystallite within a certain time interval.

The following points are crucial for the model (some of them, , i.e., items 1, 2, 4, describe mathematically a probability space (Ω,U, P) of events for the crystallites of a PC sample under PD):

1. The spectrum of mechanical energy for each crystallite at a PD value *ε* consists of discrete levels, $E_d^0$, $E_d^1$, $E_d^2$,…, $E_d^n$, …, $E_d^N$, …, depending on *ε* and starting with the lowest energy level $E_d^0$ of an ideal crystal, followed by the levels $E_d^1$ for a crystallite with a unit dislocation, and then by the levels $E_d^2$, with a dislocation of (axis) length $2b$,…, $E_d^n$, the dislocation of length $L=nb$,…, and also with $E_d^N$, being the energy of the maximum rectilinear dislocation (6). With each elementary PD act, the crystallite either acquires or loses (local restoration of crystallinity) a 1D defect with *n* atoms in its axis with the energy values $E_d^n$, for *n*=0,1,…,*N*,…. (see Fig. 4a). For each PD act, it is possible to expect the appearance of (curvilinear) dislocations with a large number $\hat{N}$ of atoms on the axis $N_d/2 > \hat{N} > N$, with $N_d$, being the number of crystallite atoms;

2. At an arbitrary time instant $t$ each crystallite may be in a state with $m_1$ unit dislocations, with $m_2$ dislocations having 3 atoms on the axis (i.e. consisting from 2 segments); ..., with $m_n$ dislocations having ($n+1$) atoms on the axis (i.e. consisting from $n$ segments) ..., with $m_N$ maximum rectilinear dislocations for $(0,...0) \leq (m_1, m_2, ..., m_n, ..., m_N)$, with the mechanical energy $\sum_{n=1}^{N} m_n E_d^n$ of all dislocations and no allowance made for the energy of elastic deformation.

3. *The minimal time* $\Delta t_0$ between PD acts, under which the crystallite is extended by the value $d(1+\varepsilon)\Delta\varepsilon = b_\varepsilon$ is connected with (properly *elementary PD act*), i.e. the appearance of two dislocations (with the *effective Burgers vectors* $b_\varepsilon$ and $-b_\varepsilon$) located in the crystallographic slip plane passing through the center of a crystallite, for a given value $\varepsilon$ of residual deformation. Due to PD homogeneity, we assume $b_\varepsilon = b(1+\varepsilon)$, whereas by the time instant $t = (\varepsilon - \varepsilon_{0.05})/\dot{\varepsilon}$ the interval $\Delta t_0$ is determined by the condition (for stretching along the $z$-axis, with $\varepsilon = u_{33}$):

$$\varepsilon - \varepsilon_{0.05} + \Delta\varepsilon = \dot{\varepsilon}t + \dot{\varepsilon}\Delta t_0 \Rightarrow \Delta t_0 = \frac{\Delta\varepsilon}{\dot{\varepsilon}} = \frac{b_\varepsilon}{\dot{\varepsilon}d(1+\varepsilon)} = \frac{b}{\dot{\varepsilon}d} \qquad (7)$$

One of such tube along the $z$ axis with the cross-sectional area $(b_\varepsilon)^2$ in a given plane is sufficient to deform the crystallite by a measurable value $\Delta\varepsilon$; however, because of the "explosive" nature of dislocation formation, such are virtually all the tubes in the slip plane, where the GD (nanopore and then dislocation) is generated. In view of polyhedral nature of the crystallite, there may be several closely situated tubes (then different atomic planes with GD) in the neighboring slip planes being parallel to the one under consideration. Deformations with the value $\Delta\varepsilon$ occur in almost all of the crystallographic planes of the crystallite spaced by the distance $nb$, $n = 1,..., \left[\frac{d}{nb}\right]$, albeit with different time intervals $\Delta t_n = b/\dot{\varepsilon}(\sqrt{d^2 - 4n^2b^2}) > \Delta t_0$. The minimal number of dislocations $N_0$ that arise during the time $t$ in order to achieve the residual PD value $\varepsilon$ is determined by the relation (taking into account their emergence in pairs):

$$N_0 = 2m_0 t/\Delta t_0 = 2m_0 \varepsilon d/b. \qquad (8)$$

In (8), $m_0$ is a *polyhedral parameter* taking into account the number of planes contributing to the deformation of a crystallite due to its polyhedral character (Fig. 4b). Note that it is not any elementary PD act that is accompanied by an emission of dislocations. Sometimes it is even a pair of dislocations formed in the previous PD act and diverging in the slip plane by the value $b_\varepsilon$ (thereby implementing the case of mobile dislocations) that is sufficient to restore the CL translational symmetry in the vicinity of these GDs, due to the mutual attraction of the nearest parallel planes (Fig. 2).

The distribution isotropy of crystallites implies for a cubic CL that the distribution of crystallographic slip planes relative to the loading $z$-axis inside the angle $\left[-\frac{\pi}{4}, \frac{\pi}{4}\right]$ is such that the minimum average value of the number of dislocations $\bar{N}_0$ for an arbitrary grain equals to $N_0/\sqrt{2}$ The case of anisotropic distributions of crystallites (textures) makes us introduce a texture factor $K = K(x, y, z)$ when calculating the number $\bar{N}_0 = \langle KN_0\rangle_V$ of averaging over all the crystallites configurations in a PC sample.

4. Let us determine the probability for any of the possible defects in an elementary PD act to occur at the time instant $t = (\varepsilon - \varepsilon_{0.05})/\dot{\varepsilon}$, as we examine the state of thermodynamic quasi-equilibrium[5] of a crystallite (with a fixed external loading) corresponding to the equidistant crystallite spectrum with a step equal to the energy of a unit dislocation for the residual plastic deformation $\varepsilon = \dot{\varepsilon}t + \varepsilon_{0.05}$ in accordance with the Boltzmann distribution:

$$\Delta E_{n+1,n}(\varepsilon) = E_d^{n+1}(\varepsilon) - E_d^n(\varepsilon) = \tfrac{1}{2}Gb_\varepsilon^3, \quad \forall n = 0,1,...,N = [d/b],..., \tag{9}$$

$$P(E_n,\varepsilon) \equiv P_n(\varepsilon) = A(\varepsilon)e^{-\frac{\frac{1}{2}Gb_\varepsilon^3}{k_BT}\frac{E_n}{E_N}} = A(\varepsilon)e^{-\frac{\frac{n}{2}Gb_\varepsilon^3}{Nk_BT}},\; n = 0,...,N,..., \; A(\varepsilon) = \frac{e^x - 1}{e^x},\; x = \frac{\frac{1}{2}Gb_\varepsilon^3}{k_BT}\frac{b}{d}, \tag{10}$$

where account is taken of $E_d^n(\varepsilon)/E_d^N(\varepsilon) \equiv E_n(\varepsilon)/E_N(\varepsilon) = E_n(0)/E_N(0)$ and $E_n(0) \equiv E_n$. The distribution (10) can be obtained using a quasi-particle interpretation of crystallite energy (see Sec.5).

---

[5] In general, a PD process in a crystallite and a PC aggregate is a non-equilibrium one, due to a change of ε, since $P(E_n, \varepsilon_1) > P(E_n, \varepsilon_2)$ for $\varepsilon_1 > \varepsilon_2$. However, the quasi-statics of external loading allows one to present a PD process as a sequence of equilibrium processes changing (skipping from one to another) at a change of ε if the relax-

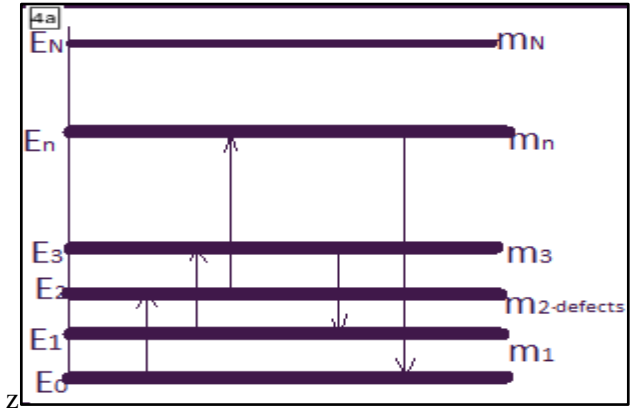

Fig. 4a : An equidistant crystallite energy spectrum distant far from the crystallite boundary (energy gaps near the boundary should be narrower) with the levels to be illustrated by a finite width, due to the thermal oscillations of the atoms. An arrow pointed from $E_m$ to $E_n, E_n > E_m$ ($E_m > E_n$) at a given instant $t$ shows a transition with an increasing (decreasing) of a crystallite energy with modifying the defect structure according to (11) inside a segment of equilibrium process.

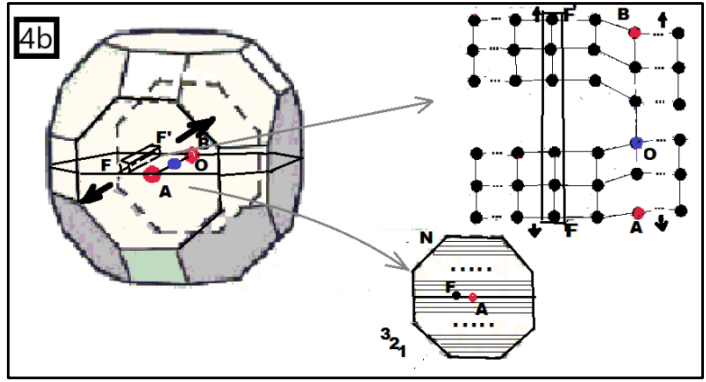

Fig. 4b The specification of the *polyhedral parameter* $m_0 = m_0(N)$ with a number $N$ (unrelated to energy level *N)* of identical parallel slip planes coincident with the direction of loading (short black arrows) and separated from each other by *b*. The crystallographic plane FAF' contains the axis of the maximal straight (rectilinearr) dislocation FA, which coincides with the thickened central line.

We assume items 1, 2, 4 to take place at a quasi-static PD with $t \geq t_1$. The space $\Omega = \{ E_d^0, E_d^1, E_d^2,..., E_d^n,..., E_d^N,...\}$ of elementary events[6] is defined for every crystallite in a state of thermodynamic quasi-

---

ation time τ for the crystallite atoms after the PD act in a stable position (a new position in the CL) is much less than the minimal time between neighboring PD acts. A natural estimation for τ is τ=$a/v_s$ = (0.3*10$^{-9}$)/10$^3$∼10$^{-12}$sec, as compared to $\Delta t_0$ =2.47*10$^{-1}$sec with the strain rate $\dot{\varepsilon}$=10$^{-5}$ c$^{-1}$, ensures the correctness of the choice for probability distribution (10) according to Boltzmann for every crystallite with a fixed ε. Under a high rate of loading, $\dot{\varepsilon}$=10$^5$–10$^8$ sec$^{-1}$, the condition $\Delta t_0$>>τ does not hold, so that the representation (10) is invalid. Thus, the probability distribution $P(E_n, \varepsilon)$ for any possible defects of an elementary PD act in a crystallite has a smooth dependence on the strain ε, $P(E_n, \varepsilon) = f_n(\varepsilon)P(E_n, 0)$, so that the quantities $f_n(\varepsilon) = [A(\varepsilon)/A(0)] (P(E_n,0)/A(0))^{\varepsilon(3+3\varepsilon+\varepsilon^2)}$ are the non-decreasing functions with $P(E_n,\varepsilon_1) > P(E_n,\varepsilon_2)$ for $\varepsilon_1 > \varepsilon_2$. Note we have chosen the factors $f_n(\varepsilon)$ in (10) as a natural multiplicative scaling of the probabilities $P(E_n, 0)$. In general, dependence of $P(E_n, \varepsilon)$ on $\varepsilon$ may be arbitrary.

[6]Assuming that in the course of an elementary PD act a dislocation may arise with its axis containing a larger quantity of atoms (segments) than the one contained at the maximal dislocation (*n>N*), the crystallite energy spectrum is augmented from above. When dislocations arise with different admissible Burgers vectors for a given CL, instead of discrete levels, the crystallite energy spectrum should consist of discrete bands, $E_d^0, E_d^{1k}, E_d^{2k},...,E_d^{nk}$,

equilibrium and is described by the occupation numbers $\vec{N} = (m_1, m_2, ..., m_n, ..., m_N)$ of the corresponding defects, as well as in terms of the probabilities of elementary events (10), which depend on the residual PD *via the effective energy* $E_d^{L_e}(\varepsilon) = \frac{1}{2}Gb_\varepsilon^3$, and thereby also on the time instant $t$. At small PD values, $\dot{\varepsilon}t = \varepsilon$, the factor $M(\varepsilon) = \frac{1}{2}Gb_\varepsilon^3/(k_BT) \approx \frac{1}{2}Gb^3/(k_BT) = M(0)$ that determines the energy scale of dislocation emergence is the inverse of speed sensitivity [23,24]. The energy value of an elastic deformation with $\varepsilon_{0.05} = u_{33}$ exhibits a cubic dependence on the crystallite size $d$. Consequently, in α-Fe, we have $F_{gr}(Fe) = 2G\frac{\mu(2\mu^2+1)}{1-2\mu}\varepsilon_{0.05}^2 \cdot \frac{m_a}{\bar{\rho}} = 0.25 \cdot 10^{-5} \text{eV} \sim 10^{-4} k_B T$ per atom, at $(\mu; m_a; \bar{\rho}) = (0.29; 9.3 \cdot 10^{-26} \text{kg}; 7800 \text{ kg/m}^3)$. The minimal time intervals between PD acts that are necessary to generate a unit dislocation (starting from nanopore) for α-Fe in CG ($d_1=10^{-4}$m) and NC ($d_2=10^{-7}$m) samples at $\dot{\varepsilon} = 10^{-5} \text{s}^{-1}$ according to (7) are equal to $(\Delta t_{01}; \Delta t_{02}) = (2.47; 2.47 \cdot 10^3) \, 10^{-1} \text{s}^{-1}$. The latter corresponds to a small defective structure of crystallites in NC materials as compared to CG materials with equal PD $\varepsilon$. In addition to the scale factor presented in the probability definition (10), there is an implicit influence of the grain boundary through the energy $E_N(d) \equiv E_d^N$ of maximal dislocation.

The transition of a crystallite from a state with energy $E_{\vec{N}_1} = \sum_{n=1}^{N} m_{1n} E_d^n$ at PD $\varepsilon$ to a state with energy $E_{\vec{N}_2} = \sum_{n=1}^{N} m_{2n} E_d^n$ at PD $\varepsilon + \Delta\varepsilon$ (for $\vec{N}_2 > \vec{N}_1$ to be lexicographically ordered) is implemented due to the absorption by the crystallite of the energy supplied by external mechanical loading with energy $M\Delta\varepsilon$ at an elementary PD act when the PC sample is lengthened on $\Delta\varepsilon = b/d$:

$$E_{\vec{N}_2}(\varepsilon + \Delta\varepsilon) = E_{\vec{N}_1}(\varepsilon) + M\Delta\varepsilon \; : \; \Delta E_{\vec{N}_1 \vec{N}_2}(\varepsilon) = E_{\vec{N}_2}(\varepsilon) - E_{\vec{N}_1}(\varepsilon) = \sum_{n=1}^{N}(m_{2n} - m_{1n})\tfrac{1}{2}nGb_\varepsilon^3$$
$$\frac{E_{\vec{N}_2}(\varepsilon+\Delta\varepsilon) - E_{\vec{N}_2}(\varepsilon)}{\Delta\varepsilon} = M + o(\Delta\varepsilon) \tag{11}$$

which describes the conservation law for mechanical energy at an set of elementary PD acts, thereby providing the changing of the strain from $\varepsilon$ to $\varepsilon + \Delta\varepsilon$.

A crystallite may emit and absorb dislocations and 0D defects under PD, thus realizing the principle of dynamic equilibrium in the form of a constant exchange of quanta $E_d^{L_e}(\varepsilon)$ between the field of mechanical (internal) stress and the crystallite. Between the external field of mechanical loading and the crystallite in a state of thermodynamic quasi-equilibrium are in a (rather one-way) process of exchanging PD energy. If a process of local CL restoration takes place, as a result of transition from a state with energy numbers $E_{\vec{N}_2}$ to a state with $E_{\vec{N}_1}$ ($\vec{N}_2 > \vec{N}_1$), then a quantum (sum of quanta) of quasi-elastic dislocation energy (a quasiparticle tentatively called a *dislocon*) is released, which can:
1) determine a new value of internal stress;
2) contribute to the growth of temperature $T$ in a crystallite;
3) be transferred to a neighboring crystallite upon interaction across the grain boundary (GB).

Let us obtain statistically a scalar dislocation density $\rho = \rho(b_\varepsilon, d, T)$, being the sum of all dislocations, both mobile and immobile ("forest" dislocations) and having different signs of Burgers vectors, $\rho = \rho_+ + \rho_-$. To this end, we calculate the average energy $\langle E_d(\varepsilon) \rangle$ of a dislocation and the number $\langle n_d(\varepsilon) \rangle$ of atoms (segments) on its axis (see Footnote 3 for partial dislocation), following the rule of averaging in ensembles, according to (10),

$$\langle E_d(\varepsilon) \rangle = A(\varepsilon) \sum_{n=0}^{\frac{N_d}{2}} E_n(\varepsilon) e^{-M(\varepsilon)\frac{E_n}{E_N}} = \tfrac{1}{2}Gb_\varepsilon^3 \left(e^{M(\varepsilon)b/d} - 1\right)^{-1}, \; \text{with} \; M(\varepsilon) = \frac{\tfrac{1}{2}Gb_\varepsilon^3}{k_B T}, \tag{12}$$

$$\langle n_d(\varepsilon) \rangle = A(\varepsilon) \sum_{n=0}^{\frac{N_d}{2}} n e^{-M(\varepsilon)\frac{E_n}{E_N}} = \left(e^{M(\varepsilon)b/d} - 1\right)^{-1} \equiv f_{N_d}(b_\varepsilon, d, T), \tag{13}$$

---

..., $E_d^{Nk}$, parameterized by the number 1, ..., $n$ of atoms in the axes, and by the number $k$ of different vectors $b_k$, $\forall n$ .The zones $E_d^{n_1 k}$, $E_d^{n_2 k}$ at $n_1 \neq n_2$ may intersect. The dependence $b_\varepsilon = f(\varepsilon)$ then implies that $E_d^n = E_d^n(\varepsilon)$.

where the factor $\frac{1}{2\pi K}\ln\left(\frac{R}{r_0}+Z\right)$ in $E_n(\varepsilon)=\frac{n}{2}Gb_\varepsilon^3$ has been omitted, and $\langle n_d(\varepsilon)\rangle$ is identical with the probability distribution function for the occurrence of a dislocation with energy $E_n(\varepsilon)$ in a grain at a state of equilibrium when the PD is equal to ε.

In the limit $d=Nb\gg b$ for CG materials, SMC and NC materials with a finite value $N\sim 10^2-10^3$, and also for grain diameters $d\leq 5$ nm, the function $f_{N_d}(b_\varepsilon,d,T)$ obeys the relations

$$\left\{\lim_{N\to\infty},\lim_{N\sim M(\varepsilon)},\lim_{N\ll M(\varepsilon)}\right\}f_{N_d}(b_\varepsilon,d,T)=\left\{M^{-1}(\varepsilon)d/b,\left(e^{M(\varepsilon)b/d}-1\right)^{-1},e^{-M(\varepsilon)b/d}\right\}, \quad (14)$$

The corresponding average dislocation energy values for CG, SMC and NC materials,

$$\left\{\lim_{N\to\infty},\lim_{N\sim M(\varepsilon)},\lim_{N\ll M(\varepsilon)}\right\}\langle E_d(\varepsilon)\rangle=\frac{1}{2}\left\{\frac{2d}{b}k_BT=\frac{N}{M(\varepsilon)}Gb_\varepsilon^3,Gb_\varepsilon^3\left(e^{M(\varepsilon)b/d}-1\right)^{-1},Gb_\varepsilon^3 e^{-M(\varepsilon)b/d}\right\}, \quad (15)$$

imply that the first value is equal to the thermal energy of $N$ atoms, being the energy of a dislocation with $NM^{-1}(\varepsilon)$ ($\sim 10^{-2}N$) atoms on its axis, i.e., basically the dislocation is adjacent to the GB from inside; the second value with energy $E_d^{L_e}<\langle E_d\rangle<10E_d^{L_e}$ describes the fact of "germination" of a dislocation in crystallites of SMC and NC materials in the form of incomplete dislocations, as well as dislocations terminating at other dislocations, and the third value at $\frac{M(\varepsilon)}{(\ln 2)}b>d$, due to $f_{N_d}<1$, implying that $\langle E_d(\varepsilon)\rangle\big|_{N<M(\varepsilon)\ln^{-1}2}<\frac{1}{2}Gb_\varepsilon^3$, corresponds to the absence (on the average) in such crystallites of dislocation emergence and also of 0D defects, which leads to the softening (unhardening) of a sample. In NC materials for all values $\varepsilon,\varepsilon<\varepsilon_0$ and a certain PD value $\varepsilon_0$, $\langle E_d(\varepsilon)\rangle\big|_{\varepsilon<\varepsilon_0}\geq\frac{1}{2}Gb_\varepsilon^3$, hardening may occur, whereas at $\varepsilon\geq\varepsilon_0$ for $\langle E_d(\varepsilon)\rangle\big|_{\varepsilon\geq\varepsilon_0}<\frac{1}{2}Gb_\varepsilon^3$ softening may take place at a quasi-static PD. The reason for the latter is the fact that there no sufficient number of the atoms in such crystallite to produce nanopores (and thereby dislocations) within thermal-fluctuation mechanism.

The length of an average dislocation, $\langle L_d(\varepsilon)\rangle=b_\varepsilon f_{N_d}(b_\varepsilon,d,T)$, and the sum of the lengths of all dislocations in an arbitrary crystallite with an accumulated PD ε, according to (8),

$$L_\Sigma(\varepsilon)=N_0\cdot\langle L_d(\varepsilon)\rangle/\sqrt{2}=\sqrt{2}\,m_0\varepsilon d\,(1+\varepsilon)\,f_{N_d}(b_\varepsilon,d,T), \quad (16)$$

determine the equilibrium (i.e. *in the absence, prior to a quasi-static PD, of a gradient due to internal stress in the grains as e.g. in annealed PC materials*) scalar density of dislocations, $\rho(b_\varepsilon,d,T)$, at the crystalline phase in a sample for approximation: $V(\varepsilon)=\frac{1}{6}\pi d^3(1+\varepsilon)[1-(1/4)\varepsilon(4-\varepsilon)]=V(0)+o(\varepsilon)^7$:

$$\rho(b_\varepsilon,d,T)=B\frac{L_\Sigma(\varepsilon)}{V(\varepsilon)}=B\frac{6\sqrt{2}}{\pi}\frac{m_0}{d^2}\varepsilon\left(e^{M(\varepsilon)b/d}-1\right)^{-1}+o(\varepsilon^2), \quad (17)$$

with allowance for a change in the grain volume (for number of materials) under PD, and with a certain constant $B$ to be chosen using the condition that (in the CG limit $d\gg b$ and in the absence of PDs, ε=0), we have $B\left(e^{M(\varepsilon)b/d}-1\right)^{-1}=d/b$, which determines the value $B=M(0)$. In the limits of CG and NC aggregates for small PDs, the value of $\rho(b_\varepsilon,d,T)$ is estimated as

$$\left\{\lim_{d\gg b},\lim_{N\sim M(\varepsilon)}\right\}\rho(b_\varepsilon,d,T)=\frac{6\sqrt{2}}{\pi}\frac{\varepsilon m_0}{bd}\left\{(1+\varepsilon)^{-3},\frac{Gb_\varepsilon^3}{2k_BT}\left(e^{M(\varepsilon)b/d}-1\right)^{-1}\right\}\sim m_0\{10^{10},10^{13}\}\,\text{м}^{-2}, \quad (18)$$

---

[7] As compared to the initial paper [38], where the average size $d_\varepsilon$ of the grains was ε-dependent similar $b_\varepsilon:d_\varepsilon=d(1+\varepsilon)$, we ignore the change of an average value of diameter $d$ under PDs, whereas for the volume we naturally suppose that along the direction of a loading axis the respective length changes by the factor $(1+\varepsilon):d\to d(1+\varepsilon)$, whereas in the other perpendicular directions the factor $(1-1/2\varepsilon)$ should be taken into account, $d\to d(1-\frac{1}{2}\varepsilon)$, so that the average volume $V(\varepsilon)$, instead of that of a 3D-sphere, is that of a rotation ellipsoid, $V(\varepsilon)=V+o(\varepsilon)$, with $V(0)=V=1/6\pi d^3$ for any grain.

which holds true [5] for experimentally observed dislocation densities under $m_0 \sim 10^1 - 10^2$ and has the form of scalar dislocation density in the Conrad model at the CG limit[8].

### 4. Generalized flow stress law and generalized Hall–Petch law for yield strength

Following [23], we suppose that Taylor's deformation (dislocation) hardening law [25], which holds true in the region of CG materials, due to the interaction energy of dislocations for tangential flow stress, $\tau \sim Gbl^{-1} \cong Gb\sqrt{\rho}$, is also valid both for NC materials and for ε>0,002:

$$\tau = \tau_f + \alpha Gb\sqrt{\rho}, \tag{19}$$

with a temperature value $T$ and a dislocation interaction constant $\alpha$, varying for different materials in the range (0.1–0.4), for frictional stress $\tau_f$ at the interaction of propagating dislocations with lattice defects and obstacles of non-deformation origin. Taking into account that the FS of a polycrystalline sample, $\sigma(\varepsilon)$, is proportional to $\tau$, $\sigma(\varepsilon) = m\tau$, $m$=3.05, according to (17) and (19), we obtain

$$\sigma(\varepsilon) = \sigma_0(\varepsilon) + \alpha m \frac{Gb}{d} \sqrt{\frac{6\sqrt{2}}{\pi} m_0 \varepsilon \frac{Gb^3}{2k_BT}} \left(e^{M(\varepsilon)b/d} - 1\right)^{-\frac{1}{2}}, \quad \sigma_0 = m\tau_f. \tag{20}$$

Expression (20) provides – with accuracy up to higher orders in the quantity $\varepsilon$ to arise in the factor at the exponent due to (17) –the main analytical result of applying of our statistical model to the determination of FS at the crystalline phase of a equilibrium single-mode PC aggregate in all grain ranges from CG to NC. This result is applicable, as we show in Sec.8, to the stages of parabolic and linear hardening down to the stages of pre-fracture and destruction. Notice the parameter values $\sigma_0(0) = \sigma(0) = 0$ and $\sigma_0(0,002) = \sigma_0$ given for $\varepsilon = 0,002$ in (1). The dependence $\sigma(\varepsilon)$ determines the FS maximum $\sigma_m(\varepsilon) = \sigma(\varepsilon)|_{d=d_0}$ as depending on the extreme grain size $d_0$. Based on a transcendental equation implied by $\partial\sigma(\varepsilon)/\partial d = 0$,

$$Q(\varepsilon)/d^2 \left(e^x - 1\right)^{-\frac{3}{2}} \left[e^x - 1 - \tfrac{1}{2}xe^x\right] = 0, \quad \text{for } x = M(\varepsilon)b/d, \tag{21}$$

for a certain $Q(\varepsilon)$ independent of $d$, the value $d_0$ can be determined numerically with accuracy up to five digits, $x = 1.59363$, and therefore:

$$d_0(\varepsilon, T) = b\frac{Gb^3(1+\varepsilon)^3}{2\cdot 1.59363 \cdot k_BT}. \tag{22}$$

The FS maximum $\sigma_m(\varepsilon)$ for a polycrystalline aggregate without a second (soft) phase is calculated as

$$\sigma_m(\varepsilon) = \sigma_0(\varepsilon) + \alpha mG\sqrt{\frac{6\sqrt{2}}{\pi}m_0\varepsilon\frac{b\cdot 1.59363}{d_0(\varepsilon,T)(1+\varepsilon)^3}} \left(e^{1.59363} - 1\right)^{-\frac{1}{2}} = \sigma_0(\varepsilon) + K(\varepsilon)d_0^{-1/2}, \tag{23}$$

with a consequent restoration of the standard Hall–Petch relation (1) for any ε, due to the flow and parabolic hardening regions, albeit with a different coefficient $K$, $k \neq K$.

For CG materials, the normal Hall–Petch law for FS at $\varepsilon = 0.002$ implies a relation between the Hall–Petch coefficient $k(\varepsilon)$ and the polyhedral parameter $m_0$:

---

[8] The analytical representation (17) and limiting cases (18) for $\rho(b_\varepsilon, d, T)$ allow for a qualitative evaluation of possible dislocation substructures (DSS), which arise in a PC aggregate at an accumulation of PDs in view of a changing CL curvature [11]. In particular, when a cellular or cellular-mesh DSS arises (oriented or non-oriented), the size of a dislocation cell is proportional to $\Lambda \sim \sqrt{d}$, according to [Conrad H. Fenerstein S., Rice L., Mater. Sci. Eng.2, 3, 157, (1967); Bay B., Hansen N., Huges D.A., Kuhlmann-Wilsdorf D. Acta Met. Mater. 40, 2, 205 (1992)]. Taking the estimation $\Lambda \sim 1/\sqrt{\rho}$, into account, it follows from (18) in CG and NC regions, that in the first case the asymptotic $\Lambda \sim \sqrt{bd(1+\varepsilon)^3/\varepsilon}$ - holds true, whereas in the second case the cellular DSS disappears. With the growth of $\varepsilon$, the value of $\Lambda$ in the CG region decreases. To be more exact, the DSS patterns should deductable from a yet unknown system of equations in partial derivatives, one of which is expected to be of diffusion type for a function $\rho(b,d,T,x,y,z,t)$.

$$\sigma(\varepsilon)|_{d>>b} = \sigma_0(\varepsilon) + k(\varepsilon)d^{-\frac{1}{2}}, \quad k(\varepsilon) = \alpha mG\sqrt{\frac{6\sqrt{2}}{\pi}m_0\varepsilon b\frac{M(0)}{M(\varepsilon)}} \Rightarrow m_0 = \frac{\pi}{6\sqrt{2}}\frac{k^2(\varepsilon)}{(\alpha mG)^2 \varepsilon b}\frac{M(\varepsilon)}{M(0)}. \quad (24)$$

This correspondence allows us to find an explicit connection between the theoretical and empirical Hall-Petch relations for a number of materials, and to discover a *temperature-dimension effect* in Secs. 6, 7. For now, we turn to a very interesting and direct consequence of crystallite energy quantization.

### 5. Quasi-particle interpretation of crystallite energy quantization under plastic deformations

Following the wave-particle duality by Louis de Broglie, a PD "*dislocon*" of energy $E_d^{L_e}$ should possess the properties of both waves and particles. This means that the energy (a quantum of PD energy) $\hbar\omega = E_d^{L_e} + W$ required to create (or change) a single 0D- or 1D- defect cannot be less than $\hbar\omega_{red} = E_d^{L_e}$, thus determining the "red" border of frequency. For instance, in α-Fe the latter equal to $\omega_{red}$(α-Fe)=5.99·10$^{15}$s$^{-1}$ for ε=0, so that *dislocons* at lower frequencies in α-Fe do not emerge, thereby failing to generate (or change) any GD in the crystallite. Since, a *dislocon* plays the role of a carrier of interaction between the CL and dislocations in a crystallite, we choose such a dispersion law that the dependence of frequency (momentum *p*) on a wave vector $k$, $\omega = \omega(k)$ is linear, as in the case of massless particles subject to the Bose–Einstein statistic (such as acoustic phonons in the Debye approximation):

$$\omega = v_d k \ (p = \hbar k); \ k_{\max} = \pi/b_\varepsilon \Rightarrow v_d = \omega/k_{\max} = \omega b_\varepsilon/\pi, \quad (25)$$

with the propagation velocity $v_d$ for a dislocon in a medium (evaluated for α-Fe) being $v_d$(α-Fe)=5.18·10$^5$ *m/s* for ε=0 (border of the first Brillouin zone). The value of $v_d$ must be larger by 2 orders of magnitude than the speed of sound ($v_s$=5.93·10$^3$м/с), so as to be related to harmonic (phonon) oscillations of a CL, albeit unrelated to a local destruction of the latter. However, as we shall see later on, the choice $v_d = M(0)v_s$ provides the correctness of a *dislocon* interpretation as that of a composite quasi-particle consisting of acoustic phonons and created at the breaking of interatomic bonds (e.g., between *A* and *A'*, see Fig.2a, at the emergence of a nanopore).

At a thermodynamic quasi-equilibrium for a fixed value ε of residual PD in each crystallite (see, footnote 5) the quasi-equilibrium process of emission and absorption of dislocons becomes established, implying that the crystallite can accommodate standing waves composed of acoustic phonons. The relation (25) means that a plane wave in a crystallite propagates in the crystallographic plane along the *z*-axis,

$$u(r) = u_0 \exp[i(\omega t + kz)], \quad (26)$$

with the periodic boundary conditions $\exp[ikz] = \exp[ik(z+d)]$ for $k\frac{2\pi n}{d}, n \in Z$. Hence, along the extension $L_\varepsilon = 2\pi/(\varepsilon d(1+\varepsilon))$ of the deformable part of the volume in the *k*-space, there is a single admissible value of *k* and the number of modes in the *k*-length units equals to $L_\varepsilon/2\pi$. The total number of modes contained inside a thin ring between the radius values *k* and (*k+dk*), with allowance for the pairing of dislocations and for the polyhedral parameter $m_0$, is evaluated as

$$dn = 2m_0 \cdot 2\pi[k + dk - k]\frac{\varepsilon d(1+\varepsilon)}{2\pi} = \frac{2m_0\,\varepsilon d(1+\varepsilon)}{v_s}d\omega = \frac{2m_0\,M(0)\varepsilon d(1+\varepsilon)}{v_d}d\omega, \quad (27)$$

taking into account the fact that dislocons (phonons) are localized entirely in the zone of localized deformation, i.e., in the plastically deformed part of linear volume ($L = \varepsilon d(1+\varepsilon)$). Those of them which have the frequencies $\omega \geq \omega_{red}$ are the only ones that permits a dislocation (nanopore) to be created. The bandwidth of the frequencies for such dislocons should be quite narrow, $(\varepsilon) - \omega_{red} \ll \omega_{red}$, so that due to Einstein's proposal we naturally assume for the dislocons that all of their frequencies should be identical to $\omega_{red}$, which corresponds to the insertion of a Dirac δ-function, $\pi^{-1}\omega \cdot \delta(\omega - \omega_{red})$, into (27) as integration over frequencies is carried out:

$$N^* = \int_0^\infty \frac{2\,m_0\varepsilon d(1+\varepsilon)}{\pi v_s}\omega \cdot \delta(\omega - \omega_{red})d\omega = \frac{2m_0 M(0)\omega_{red}\,\varepsilon d(1+\varepsilon)}{\pi v_d} = \frac{2m_0 M(0)\varepsilon d(1+\varepsilon)}{b(1+\varepsilon)}. \quad (28)$$

Making allowance for crystallite distribution in a PC sample to de isotropic leads to the coincidence of (28) with the number of dislocations $N^*/\sqrt{2} = N_0/\sqrt{2}$ for the PD value ε obtained earlier from mechanical reasons in Eq. (8) of Sec. 3, with account taken of inserting the constant, $B = M(0)$ into (17).

The quasi-particle interpretation allows one to justify the distribution (10) of energy states in crystallites, $\{P(E_n, \varepsilon)\}$. Indeed, considering an assembly of dislocons as a gas, we can approximately regard it to be weakly interacting. At an instant $t$ when the value of PD equals to $\varepsilon$ the gas pressure before and after a PD act in a crystallite takes the values $p(\varepsilon)$ and $p(\varepsilon + \Delta\varepsilon) < p(\varepsilon)$. The energy released in the crystallite reads

$$[p(\varepsilon + \Delta\varepsilon) - p(\varepsilon)]V = -N(\varepsilon)\Delta E, \qquad (29)$$

where $N(\varepsilon)$ is the number of dislocons in the crystallite volume $V$. A state equation for this gas can be written at an instant $t$, with allowance for the fact that a 1D-defect contains $N = \beta[d/b]$ atoms (for $0 < \beta \leq 1$) in its axis, with the dislocations not necessarily passing through the diameter, where $\beta = 1$,

$$p(\varepsilon)V = N(\varepsilon) \cdot N \cdot k_B T \Rightarrow \Delta p(\varepsilon) = \frac{\Delta N(\varepsilon)}{b/d}\beta k_B T, \qquad (30)$$

In terms of $\Delta N(\varepsilon)$ the equation (29) acquires the form $\frac{\Delta N(\varepsilon)}{N(\varepsilon)} = -\beta^{-1}\frac{\Delta E}{k_B T}\frac{b}{d}$, and implies for the differentials $dN(\varepsilon), dE(\varepsilon)$ a solution of the corresponding differential equation with the boundary condition $\Delta E|_{\Delta\varepsilon=0} = 0$, namely,

$$N(\varepsilon + \Delta\varepsilon) = N(\varepsilon)e^{-\frac{1}{\beta k_B T}\frac{\Delta E}{d}\frac{b}{d}}, \qquad \Delta E = E(\varepsilon + \Delta\varepsilon) - E(\varepsilon). \qquad (31)$$

The distribution (31) for the number of dislocons, and for their concentration $n(\varepsilon) = \frac{N(\varepsilon)}{d^3}$, corresponds to the Boltzmann distribution employed to determine the statistical model of Sec. 3, albeit for a discrete change of $\Delta E$ with $\beta = 1$..

### 6. Generalized Hall–Petch law as implemented for α-Fe, Cu, Al, Ni, α-Ti, Zr

To construct a theoretical dependences for the HP law (20) in specific PC materials, we should determine the values of the constant $m_0$ (24). To this end, we use the experimental data of Table 1 on the HP coefficient $k(0,002)$ for PC single-mode samples with BCC, FCC and HCP CL, with the corresponding values of $\sigma_0$, $G$, lattice constant $a$ [46], Burgers vector with the least possible length $b$ for the respective most probable sliding systems (see Table 2), the constant of interaction for a dislocation $\alpha$ [23, 46] the calculated values of the least unit dislocation $E_d^{L_e}$, extreme grain size $d_0$, maximal difference of $\sigma_y$ according to (20) and (24) for $T$=300K:

| Type of CL | BCC | FCC | | | HCP | |
|---|---|---|---|---|---|---|
| Material | α-Fe | Cu | Al | Ni | α-Ti | Zr |
| $\sigma_0$, MPa | 170 (annealed) | 70 (anneal.); 380 (cold-worked) | 22 (anneal. 99,95%); 30 (99,5%) | 80 (annealed) | 100(~100%); 300 (99,6%) | 80-115 |
| $b$, nm | $\frac{\sqrt{3}}{2}a$ =0.248 | $a/\sqrt{2}$ =0.256 | $a/\sqrt{2}$ =0.286 | $a/\sqrt{2}$ =0.249 | $a$=0.295 | $a$=0.323 |
| $G$, GPa | 82.5 | 44 | 26.5 | 76 | 41.4 | 34 |
| $T$, K | 300 | 300 | 300 | 300 | 300 | 300 |
| $k$, MPa·$m^{1/2}$ | 0.55-0.65 ($10^{-5}-10^{-3}$m) | 0.25 ($10^{-4}-10^{-3}$m) | 0.15 ($10^{-4}-10^{-3}$m) | 0.28 ($10^{-5}-10^{-3}$m) | 0.38-0.43 ($10^{-5}-10^{-3}$m) | 0.26 ($10^{-5}-10^{-3}$m) |
| $\alpha$ | – | 0.38 | – | 0.35 | 0.97 | – |
| $E_d^{L_e} = \frac{1}{2}Gb^3$, eV | 3.93 | 1.28 | 1.96 | 3.72 | 3.33 | 3.57 |
| $m_0 \cdot \alpha^2$ | 3.66-5.11 | 2.57 | 2.28 | 1.11 | 5.83-7.47 | 3.69 |
| $d_0$, nm | 23.6 | 14.4 | 13.6 | 22.6 | 23.8 | 28.0 |
| $\Delta\sigma_m$, GPa | 2.29-2.56 | 1.34 | 0.83 | 1.20 | 1.58-1.70 | 1.00 |

**Table 1**: The values $\sigma_0$, $\Delta\sigma_m=(\sigma_m-\sigma_0)$, $E_d^{L_e}$, $k$, $m_0$, $\alpha$ in BCC, FCC and HCP polycrystalline metal samples with $d_0$, $b$, $G$ obtained using [46] at $\varepsilon = 0.002$.

The values for $k$ at $\varepsilon = 0.002$ are used for α-Fe, Cu, Ni [5], for Zr [46], for Al [45], and for α-Ti [47,48] with the range of grain size shown in the brackets:

| | α-Fe | Cu | Al | Ni | α-Ti | Zr |
|---|---|---|---|---|---|---|
| **Plane** | {110}, {112}, {123} | {111} | {100},{111} | {111} | $(10\bar{1}0), (10\bar{1}1), (0001)$ | $(10\bar{1}0)$ |
| **Direction** | <111> | <110> | <110> | <110> | $[2\bar{1}\bar{1}0]$ | $[11\bar{2}0]$ |

**Table 2**: The most probable sliding systems at *T*=300K [46] for α-Fe, Cu, Al, Ni, α-Ti, Zr in terms of the Miller indices for BCC, FCC and the Miller–Bravais indices for HCP lattices.

The graphic dependence $\sigma_y = \sigma_y(d^{-1/2})$ for the crystallite phase of PC aggregates of α-Fe, Cu, Al, Ni, α-Ti, Zr with closely-packed randomly oriented grains, to be homogeneous with respect to their size (single-mode case) at T=300K, are shown by Fig. 5 on the basis of Tables 1, 2.

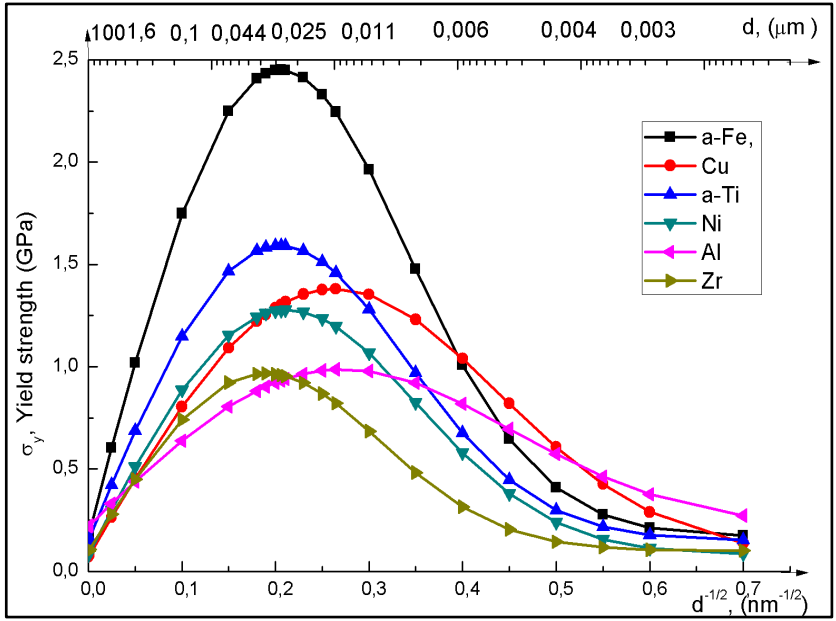

**Fig. 5.** Graphic dependence (plots) for generalized Hall-Petch law (20), (24) at $\varepsilon = 0,002$ with an additional upper scale with size of grains *d* given in μm. Upper axis *d* is changing within range (∞;0) with the inverse quadratic scale and the correspondence (100; 1,6; 0,1; 0,044; 0.025; 0,011; 0,006; 0,004; 0,003) μm ↔ (0,005; 0,015; 0,1; 0,15; 0,2; 0,3; 0,41; 0,5; 0,57) nm$^{-1/2}$ for the respective values on the lower axis.
The least possible values of the parameter $m_0(k)$ for the α-Fe, α-Ti, values of σ$_0$ for annealed materials with the maxima of $\sigma_y$ calculated for the extreme grain size values $d_0$ relative to Table 1.

According to Fig. 5, experimental data coincide approximately at the extreme size values [5], as well as the values for maximums σ$_m$ [42]. The values of $d_0(0.002, 300)$, e.g., for α-Fe, Cu, Al, Ni, α-Ti, Zr given in Table 1, is in complete agreement with the range (both empirical and theoretical) of critical size values for the average diameters of grains $d_{cr}$ for PC samples (listed, e.g., in Ref. [5] (Table 2.6, pp. 110–111) and in Ref. 56), ranging from 5–10 nm to 20–50 nm, particularly, for $d_{cr}(Cu)=10$ nm ≈ $d_0(Cu)=14.4$ nm from Ref. [21] and $d_{cr}(Ni)=20$ nm ≈ $d_0(Ni)=22.6$ nm. For the corresponding values for the maxima of experimental σ$_y$, i.e., $\bar{\sigma}_m(0.002)$ and $\sigma_m(0.002)$ in single-mode (on average) PC samples (see, e.g., Refs. [42], [57]), we find that $\bar{\sigma}_m(0.002)$ (α-Fe) ≈ 2.75 GPa, $\bar{\sigma}_m(0.002)$ (Ni) ≈ 1.7 GPa, $\bar{\sigma}_m(0.002)$ (Cu) ≈1.0 GPa coincide approximately with the theoretical maxima, with allowance for the various definitions of HP coefficients according to the literature (see Table 2 in [56] with $k(Cu) \in [0.01,0.024]$ for UFG PC samples). The difference for Ni and Cu may be explained by leaving out of account, first, unhardening due to weak grain boundary parts, which leads to a decrease in $d_0$, $\sigma_m$, and, second, excitation at PDs of other dislocation ensembles, especially in the NC region with a Burgers vector $b_1$ larger than the one for the most probable dislocation, and therefore with a larger unit dislocation energy, due to $E_d^{Le}(G,b) < E_d^{Le}(G,b_1)$, and a larger input to $\sigma(0.002)$, as for Ni. Therefore, the maximum values σ$_m$ demand taking into account a negative input from the GB phase, which is examined in forthcoming Sections 9, 10.

### 7. Temperature dependence of yield strength and extreme grain size for Al

Since the growth of temperature causes a decrease in the value of shear module $G(T)$ (as well as $\sigma_0(T)$), whereas the linear parameters *b* and *d* increase with the same true linear coefficient of the temperature expansion $\alpha_d$ [45] (for BCC and FCC materials), then the extreme grain size $d_0(\varepsilon)$ is shifted to the region of smaller grains, $d_0(\varepsilon,T) > d_0(\varepsilon,T')$ for $T' > T$; $T',T \in [T_1,T_2]$ at the same phase in a given material, according to the rule:

$$d_0(\varepsilon,T') = b_\varepsilon(T') \frac{\frac{1}{2}G(T')[b_\varepsilon(T')]^3}{1,59363 \cdot kT} = d_0(\varepsilon,T)g(\alpha_G,\alpha_d,T,T'),$$

$$g(\alpha_G,\alpha_d,T,T') = \left(\frac{b_\varepsilon(T')}{b_\varepsilon(T)}\right)^4 \frac{G(T')T}{G(T)T'} = \left(1 + \alpha_d(T'-T)\right)^4\left(1 - \alpha_G(T'-T)\right)\frac{T}{T'}, \quad (32)$$

$$for\ b_\varepsilon(T') = b_\varepsilon(T)\left(1 + \alpha_d(T'-T)\right);\quad G(T') = \left(1 - \alpha_G(T'-T)\right)G(T)$$

with a linear temperature coefficient for the shear modulus $\alpha_G$, e.g. for $Al$: $\alpha_G = 5,2 \cdot 10^{-4} K^{-1}$ being approximately constant within temperature range [250K,300K]. It follows from (32) that for $T$ varying in a small range the value $d_0(\varepsilon,T)$ changes multiplicatively with the factor $g(\alpha_G,\alpha_d,T,T')$. At the same time with an accommodation of PD $\varepsilon$ the value of $d_0(\varepsilon,T)$ is shifted to the area of larger grains: $d_0(\varepsilon_1,T) > d_0(\varepsilon_2,T)$ for $\varepsilon_1 > \varepsilon_2$. The behavior of $\sigma(\varepsilon)$ and $\sigma_m(\varepsilon)$ for a monotonic change of the temperature is composed of the $T$-behavior in a crystal and of the dislocation substructure at the crystallite phase of a sample. The quantities $(\sigma_0, G) = (\sigma_0(T), G(T))$ corresponding to the crystal substructure decrease and $(b, d)$ increase with a grows of $T$, whereas for the latter (dislocation) substructure the explicit T-dependence in the relations (20), (24) produces an increase in $\sigma$, $\sigma_m$ with grows of $T$. The literature available to date provides *no systematic experimental data* concerning the temperature dependence of $\sigma(\varepsilon)$, $\sigma_0(\varepsilon), d_0(\varepsilon)$, which makes it necessary to fill this gap. For instance, the results of molecular dynamics simulation for Cu [22] supply contradictory data to the effect that the growth of $T$ causes the values of $\sigma_y$ and the maximum $\sigma_m(\varepsilon_{0,2})$ to decrease for any fixed $d$; besides, the simulated values of $d_0(\varepsilon_{0,2})$ are shifted to the area of larger grains, from 4 nm at T=280 K to 25 at T=370K. The latter results, (obtained under high-speed mechanical loading with $\dot\varepsilon_. = 5 \cdot 10^8\ s^{-1}$) contradicts to the established displacement law (22). However, one should point out that (22) is derived assuming the existence of the probability distribution $P_n(\varepsilon) = A(\varepsilon)e^{-\frac{\frac{1}{2}Gb_\varepsilon^3}{k_BT}\frac{E_n}{E_N}}$ given by Eqs. (10) in the case of quasi-static loading at thermodynamic quasi-equilibrium for a given $\varepsilon$, which plays the role of an adiabatic parameter. For the maximal differences: $\Delta_m\sigma(\varepsilon,T) = \sigma_m(\varepsilon,T) - \sigma_0(\varepsilon,T)$ calculated at $T'>T$ we have the following ratio:

$$\frac{\Delta_m\sigma(\varepsilon,T)}{\Delta_m\sigma(\varepsilon,T')} = \frac{G(T)}{G(T')}\sqrt{\frac{b_\varepsilon(T)d_0(\varepsilon,T')}{b_\varepsilon(T')d_0(\varepsilon,T)}} = (1 + \alpha_G(T'-T))\sqrt{(1-\alpha_d(T'-T))g(\alpha_G,\alpha_d,T,T')} < 1, \quad (33)$$

An approximated right-hand side, $r(\varepsilon,T,T') = \{1 + (0,5\alpha_G + 1,5\alpha_d)(T'-T)\}\sqrt{T/T'}$, where a decreasing root $\sqrt{T/T'}$ suppresses the growth of the first multiplier, e.g., in Al at the range $[T,T'] = [300,350]K$ with $\alpha_d(Al)=2,33 \cdot 10^{-5} K^{-1}$ [45], justifies the estimate $\Delta_m\sigma(\varepsilon,T)/\Delta_m\sigma(\varepsilon,T') = 0,93$.

Experimental estimates for the growth $\sigma(\varepsilon,T) > \sigma(\varepsilon,T')$ in this case indicate that stress $\sigma_0(\varepsilon,T)$ should decrease more rapidly than $\Delta_m\sigma(\varepsilon,T)$ increases with the growth of temperature. However, the value of $\sigma_0(\varepsilon,T)$ can range from 7% to 20% of the value $\sigma_m(\varepsilon,T)$ at T=300 K in various materials (see Table 3 below). In the low-temperature region, the value of $\sigma_y(d,T)$ undergoes a significant increase, explained in [46] by a predominance of twinning in BCC, FCC and especially in HCP metal polycrystals. Note that the process of twinning does not contribute to plasticity, but rather causes an emergence of additional sliding systems for dislocations (due to a change in CL curvature-torsion). For twinning at the CG region of a PC material, a relation for $\sigma_y(d,T)$, along with $\sigma_{0tw}(T)$ and with an HP coefficient $K_{tw}(T)$, has been obtained in a form analogous to the normal HP law (24) with the first quantity being smaller than the one for pure dislocations, $\sigma_0(T)$, and with the second quantity being much larger (5 times larger [46] for Cr), than the one for $k(\varepsilon)$. The description problem for an input of twinning into deformation hardening, along with a further study of $T$-dependence for $\sigma_y(d,T)$, is beyond the scope of this paper. However, it is quite remarkable that twinning (as producing 2*D*-defects) can also be studied in terms of partial dislocation clusters![9] (see Chap. 23 in [34] for details). The condition for $\sigma(\varepsilon,T)$ to decrease almost everywhere with the growth of $T \in [T_1, T_2]$ is equivalent to the following::

$$\frac{d\sigma}{dT} = \frac{d\sigma_0}{dT} + \alpha m \frac{Gb}{2d} \left\{ \left(e^{M(\varepsilon)b/d} - 1\right) \frac{d}{dT} \ln\left(\frac{G^2}{M(0)}\right) - \frac{b}{d} e^{M(\varepsilon)b/d} \frac{dM}{dT} \right\} \sqrt{\frac{6\sqrt{2}}{\pi} m_0 \varepsilon (1+\varepsilon) M(0)} \left(e^{M(\varepsilon)b/d} - 1\right)^{-\frac{3}{2}} \leq 0. \quad (34)$$

In CG and fine-grain (FG) materials ($b \ll d$) the form of (20) is that of the normal HP law (24), which is also valid at $\varepsilon > 0{,}002$, in accordance with [23], implying that, in view of the *T*-independence of the ratio $b/d$, the growth of *T* causes the yield strength $\sigma_y(T)|_{d \gg b} = \sigma(0{,}002; T)|_{d \gg b}$ to decrease at a decrease in $\sigma_0$ and *G*. Outside the GC, FG and UFG regions, the FS has an opposite temperature behavior.

Therefore, a 3-dimensional plot $(d, T, \sigma(\varepsilon, d, T))$ of Eq. (20) contains the value $d_1(\varepsilon)$ ($d_1 \approx 3d_0 \gg b$). An estimate for $d_1(\varepsilon)$ is implied by a cubic approximation of the HP law obtained from the general form (20) by keeping the quadratic terms of $\left(e^{M(\varepsilon)b/d} - 1\right)$ in powers of $b/d$, as follows:

$$\sigma(\varepsilon)|_{d \gg b} = \sigma_0(\varepsilon) + k(\varepsilon) d^{-1/2} (1 - \tfrac{1}{4} M(\varepsilon) b/d) \implies d_1 = \frac{bM(\varepsilon)}{4\alpha_G}\left(\frac{1}{T} + \alpha_G\right). \quad (35)$$

Indeed, the dependence of $\sigma(\varepsilon)|_{d \gg b}$ (35) on *T* [as we omit $\frac{d\sigma_0}{dT}$ and keep the leading term of an explicit dependence on *T* in $M(\varepsilon)$] implies that $d_1 = \frac{bM(\varepsilon)}{4\alpha_G}\left(\frac{1}{T} + \alpha_G\right)$. For Al at *T*=300 K, we find $d_1 = 138b$ =39.6 nm. Therefore, $d_1 \approx 3d_0(\varepsilon, T)$, so that for all the values $d > d_1(\varepsilon)$ the stress $\sigma(\varepsilon, d, T)$ decreases with the grows of T, whereas for all the values $d < d_1(\varepsilon)$ the quantity $\sigma(\varepsilon, d, T)$ increases (with the subsequent maxima of $\sigma(\varepsilon, d, T)$ with respect to the average size of crystallites and the maxima relative to *T* at a fixed $\varepsilon$). For low values of *T*, the model should provide a significant increase in $\sigma(\varepsilon, d, T)$ and thereby is in need of a clarification; however, for $d \gg bM(\varepsilon)$ the value of FS is determined, due to (24), by the behavior of *G* and by the yet insufficiently studied behavior of $\sigma_0(T)$.

As in the case of *G*, we expect a similar *T*-dependence for $\sigma_0(T)$, being a characteristic of a monocrystal. Concerning the HP coefficient, we note, in the first place, that $k(\varepsilon)$ in (24) determined in the CG limit does not depend explicitly on temperature (for chromium, see Fig. 2.16 in [46] up to 100K), ); secondly, there exists as many as 10 analytical definitions of $k(0{,}002)$ in [5] (see also Table 5 in [46]).

The temperature dependence of the yield strength $\sigma_y(d,T)$ in the coordinates $(d^{-1/2}, \sigma_y)$, for instance for Al, is presented in a parametric form by Fig. 6, as we assume for the parameter $\sigma_0(T)$ to have the same dependence as that of $G(T)$, e.g., the one given by the factor $[1 - 5.2 \cdot 10^{-4}(T' - T)]$, considering the absence of experimental data.

| *T*,K \ Al | *G*, GPa | $\sigma_0$, MPa | $d_0$, nm | ($\sigma_m$-$\sigma_0$), GPa |
|---|---|---|---|---|
| 350 | 25.8 | 21 | 11.3 | 0.85 |
| 300 | 26.5 | 22 | 13.6 | 0.83 |
| 250 | 27.4 | 23 | 16.8 | 0.74 |
| 200 | 28.1 | 23.5 | 21.5 | 0.67 |
| 150 | 28.8 | 24 | 29.5 | 0.59 |

**Table 3:** The values of parameters $\sigma_0, d_0, G, (\sigma_m\text{-}\sigma_0)$ for Al in the temperature range [150,350]K.

---

[9] Analytically, it can be interpreted that the influence of twinning on deformation hardening is already taken into account due to dislocations which compose the twins..

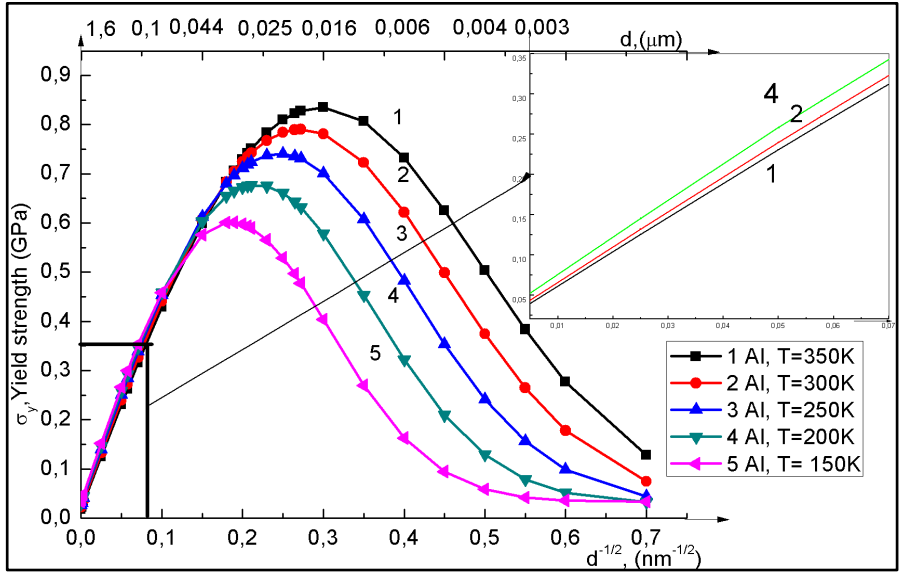

**Fig. 6.** Plotted graphic dependence for the generalized Hall–Petch law (20) at $\varepsilon = 0,002$ for *Al* at $T = 350; 300; 250; 200; 150K$. At the input, part of the dependence in the CG and FG regions is indicated (selected by a rectangle) at $T = 350K, 300K, 200K$ under the normal HP law (24). The straight lines numbered 3 for $T=250K$ and 5 for $T=150K$ are located at the input, respectively, between lines 2, 4 and above line 4.

From the Fig. 6 it follows that in the grain range up to 0,1 μm the *T*-behavior of $\sigma_y(d,T)$ is the usual one ($\sigma_y$ decreases with the grows of *T*), whereas for $d<d_1=39,6$ nm, as one goes to the anomalous part of the HP law, the *T*-behavior of $\sigma_y(d,T)$ looks unusual (as $\sigma_y$ increases with grows of *T*). In particular, for $d=d_1$ the yield strength $\sigma_y(d_1,T)$ does not depend on *T* in a wide range of temperatures. For instance, at *T*=150 K the extremal values of dislocation density and yield strength are reached at $d_0(150)=29.5$ nm, and then with a decrease in *d* (e.g., down to *d*=21 nm) the dislocations exit the grain bodies and unhardening to follow, so that $\sigma_y(29,5;150) > \sigma_y(21;150)$. At the point such a sample of Al under stress is heated up to *T*=250 K this grain of *d*=29.5 nm is already pre-extreme ($29,5>d_0(250)=16,8$ nm), , so that when *d* decreases down to *d*=21 nm, under the given plastic deformation the scalar density of dislocations increases along with the increase in yield strength, and thus the inverse inequality holds true, $\sigma_y(29,5;150) < \sigma_y(21;150)$. This kind of behavior should be expected for any PC materials with closely packed grains and is yet to be verified experimentally (with allowance for a possible change of grains due to recrystallization, especially in Al). We refer to this phenomenon as a *temperature-dimension effect* (TDE) in PC materials, which is characterized by the two following properties, at least in a sufficiently wide range of *ε* under plastic deformations:

1) a displacement of the extremal size value $d_0(\varepsilon, T)$ (22) to the large grains region with a decreasing of the temperature, $d_0(\varepsilon, T_1) > d_0(\varepsilon, T_2)$ for $T_1 < T_2$;

2) an increase of FS $\sigma(\varepsilon)$, including the maximum $\sigma_m(\varepsilon)$ (23), with a grows of T in the NC region in single-mode PC materials for $d < d_1 \approx 3d_0$, and a decrease for $d > d_1$.

An including of the second GB phase in the model may significantly change this effect as it we show in forthcoming Sec. 11.

### 8. Stress-strain curves for the crystallite phase of α-Fe. Backofen–Considére criterion

The dependence (20) of $\sigma(\varepsilon)$, together with the stress-strain curve $\sigma = \sigma(\varepsilon)$ in Fig. 7, allows one to find the strain hardening coefficient $\theta(\varepsilon) = d\sigma/d\varepsilon$, assuming that $\theta_0(\varepsilon) = d\sigma_0/d\varepsilon$,

$$\theta(\varepsilon) = \theta_0(\varepsilon) + \alpha m \frac{Gb}{d\sqrt{\varepsilon}} \sqrt{\frac{3}{\pi\sqrt{2}} m_0 M(0)} \left\{ e^{M(\varepsilon)b/d} - 1 - \frac{3\varepsilon}{1+\varepsilon} M(\varepsilon) \frac{b}{d} e^{M(\varepsilon)b/d} \right\} \left( e^{M(\varepsilon)b/d} - 1 \right)^{-\frac{3}{2}}. \quad (36)$$

The stress-strain curves for the dependence $\sigma = \sigma(\varepsilon, d, T)$ (20) at the pure crystalline phase of a PC α-Fe sample at *T*=300K for various average grain sizes are given by Fig. 7 on the basis of Table 4. The values of conditional elastic limit σ(0,0005) are formally calculated according to (20).

| α-Fe | $\sigma(\varepsilon)-\sigma_0(\varepsilon)$, GPa | | | | | | | | | |
|---|---|---|---|---|---|---|---|---|---|---|
| $\varepsilon$, x10$^{-2}$ $d, nm$ | 0,05 | 0,1 | 0,2 | 0,5 | 1 | 2 | 5 | 10 | 20 | 30 |
| 10$^6$ | 0.009 | 0.012 | 0.017 | 0.027 | 0.038 | 0.054 | 0.081 | 0.107 | 0.133 | 0.144 |
| 10$^5$ | 0.028 | 0.039 | 0.055 | 0.086 | 0.122 | 0.169 | 0.256 | 0.338 | 0.420 | 0.456 |
| 10$^3$ | 0.273 | 0.386 | 0.545 | 0,,856 | 1.204 | 1.677 | 2.536 | 3.339 | 4.129 | 4.465 |
| 150 | 0.667 | 0.943 | 1.332 | 2.095 | 2.938 | 4.086 | 6.147 | 8.016 | 9.690 | 10.197 |
| 23.6 | 1.145 | 1.617 | 2.279 | 3.572 | 4.976 | 6.827 | 9.840 | 11.859 | 11.861 | 9.834 |
| 10 | 0.827 | 1.166 | 1.640 | 2.548 | 3.500 | 4.666 | 6.140 | 6.254 | 4.192 | 2.130 |

**Table 4**: The values of stress $\sigma(\varepsilon)-\sigma_0(\varepsilon)$ for the crystalline phase of a single-mode polycrystalline α-Fe at $T=300K$ for various average grain size values the range of $\varepsilon \in [0.0005;0.3]$.

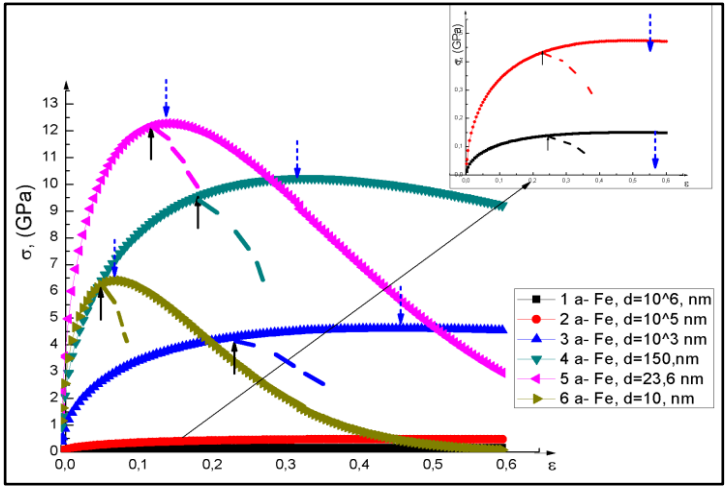

**Fig. 7**. Plots $\sigma = \sigma(\varepsilon,d,T)$ (20) for α-Fe, $m_0 \cdot \alpha^2 = 3,66$, with stress-strain curves 1, 2, 3, 4, 5, 6 for $d=10^{-3}$; $10^{-4}$; $10^{-6}$m; $d=150$ nm; $d=d_0=23,6$ nm: $d=10$ nm for $T=300K$. By the black arrows below it is indicated the values, where the Backofen-Considére condition (39) is realized with $\varepsilon_{\text{fr.cond.}}$ (40) and expected dashed lines of curves before the fracture. The blue arrows from top indicate the maximums of $\sigma(\varepsilon)$ for the true strains calculated according (38). On the input the plots for the stress-strain curves for CG aggregates are shown.

An obtaining the value $\sigma_{\max}(\varepsilon)$ follows from the condition $d\sigma/d\varepsilon = 0$, similar to the one given by (22),

$$\left(e^{y(\varepsilon)} -1\right)^{-\frac{3}{2}} \Big/ 2\sqrt{\varepsilon} \left\{ e^{y(\varepsilon)}\left(1 - \frac{3\varepsilon}{1+\varepsilon} y(\varepsilon)\right) - 1 \right\} = 0 \ \ with \ \ y(\varepsilon) = M(\varepsilon)b/d \ . \tag{37}$$

The extreme value of the PD quantity $\varepsilon_m$ under a quasi-static loading follows from an approximate solution of dimensionless $y_0(\varepsilon)$ of a transcendental equation for the expression within the figure brackets (37), which depends on the parameter $\varepsilon$,

$$y_0(\varepsilon) = \frac{Gb_\varepsilon^3}{2k_B T}\frac{b}{d} \ \Rightarrow \ \varepsilon_m(d,T) = f(y_0): \ \varepsilon_m(d,T) = \sqrt[3]{y_0(\varepsilon_m)d/(M(0)b)} - 1 \ , \tag{38}$$

where the root $\varepsilon_m$ is taken in the range $0 < \varepsilon_m < 1$. For a dependence $\sigma = \sigma(\varepsilon)$ of the form (20), the relation $\lim_{d \gg b} \varepsilon_m = 0.5$ holds true.

In constructing the stress-strain curves, we use the Backofen–Considére condition of fracture, $\sigma = d\sigma/d\varepsilon$, see [5], which selects the regions of homogeneous and localized PD, and permits us to determine the values of conditional strain $\varepsilon_{\text{fr.cond.}}$ and stress of fracture (ultimate stress) $\sigma_S$ from the equation

$$\theta_0(\varepsilon) - \sigma_0(\varepsilon) = \alpha m \frac{Gb}{d\sqrt{\varepsilon}} \sqrt{\frac{3}{\pi\sqrt{2}} m_0 M(0)} \left(e^{M(\varepsilon)b/d} - 1\right)^{-\frac{1}{2}} \times X(\varepsilon,d,T),$$

$$X(\varepsilon,d,T) = 2\varepsilon - 1 + \frac{3\varepsilon}{1+\varepsilon} M(\varepsilon) \frac{b}{d} e^{M(\varepsilon)b/d} \left(e^{M(\varepsilon)b/d} - 1\right)^{-1} , \tag{39}$$

with allowance for (36), (37). In the approximation $\theta_0(\varepsilon) = \sigma_0(\varepsilon) = 0$, the solutions of (39) $\varepsilon_{\text{fr.cond.}}(d,T)$ are determined by the condition $X(\varepsilon,d,T) = 0$. For the values $d$ in Table 4 at $\varepsilon_{\text{fr.cond.}}(d,300)$, we have

$$\varepsilon_{fr\ cond}(10; 23.6; 150; 10^3, 10^5; 10^6) = (0.065; 0.12; 0.19; 0.225; 0.23, 0.23) \tag{40}$$

The absolute maximum value $\sigma_{\max}(\varepsilon_m, d_m, 300) = 13.27$ GPa (for $\sigma_0(\varepsilon)=0$) is determined numerically from equations (20), (38), with $d_m = 40.6$ nm and $\varepsilon_m = 0.2$. This value was not experimentally observed in the NC region. It is close to the theoretical ultimate stress and is determined by the peculiarities of the model. Among the peculiarities one may select, first, a single-mode property of crystallites, second, not accounting for the second phase (using the terminology of Ref.[5]) with grains boundary as the regions between the pure crystallites, being filled by the crystallites of sizes $d_{C2} \ll d$ and by pores, considered, e.g. as accommodations of the nanopores (see Fig. 3 with composite model for crystallites labeled as *b*).

### 9. Two-phase model for polycrystalline aggregate. Unhardening due to boundary grains

To overcome the difficulties related to the above peculiarities for PC sample of α-Fe, we adopt natural assumptions, first, that an input from the second phase should be added additively in $\sigma(\varepsilon)$ (20) with a proportionality coefficient $\kappa_1, 0 \leq \kappa_1 < 1$ (the value $\kappa_1 = 0$ means the absence of an explicit contribution from the grains boundary (GB) into hardening). Second, with increasing of an accumulated PD ε, a average volume of intergrain regions is increased with changing of the second phase contents, and therefore the porous structure therein is increased, that leads on the final stage to appearance of cracks and destruction of the sample, as well as for sub-microcrystalline and NC materials this phase provides the slipping through pores of grains (or groups of grains) for sufficiently small PD, that it is by a previously unaccounted softening factor. We take into account the contribution from the GB in to the total stress of the sample by means of subtracting the stress in the porous part of the second phase of the aggregate with a coefficient of proportionality $\kappa_2, 0 \leq \kappa_2 \leq \kappa_1$. Discontinuity of the sample in areas of grain boundaries implies a necessity of a negative input from pores into total $\sigma_\Sigma(\varepsilon)$ as depending on the average size of such pores, increased by the PD accumulation, ε. Let us consider pores as *formal crystallites* of an average size $d_P$ of the same materials. Such a model of two-phase system is reminiscent of the composite models [5]. We relate the pores in a material to the GB size: considering that the larger $d_P$ (and $d_{C2}$), the larger the part of large-angle GBs, and vice-versa, the smaller $d_P$ (and $d_{C2}$), the larger the part of small-angle GBs.

One may estimate the value of a parameter $\kappa_1$ as part (weight) of a volume of the second phase of the composite model for crystallite around the, so called hard first-phase crystallite itself, with respect to the volume $V_C$ of the crystallite, so that the second-phase crystallites forms the first part of the shell (with one layer) around pure crystallite, whereas the pores part forms the second part of the shell (within the same layer) around pure crystallite plus second-phase crystallites. If the respective volumes of first-phase crystallite, second-phase crystallites and porous part from the composite model for the grain are equal to: $V_C$, $V_{C2}, V_P$ with the respective average diameters , $d_{C2}, d_P$ , so that the total volume of the second-phase is $V_{GB} = V_{C2} + V_P$, then approximating the volumes $V_C$, $(V_C + V_{C2}), (V_C + V_{GB})$ by respective 3D balls, one can estimate the weights as

$$\kappa_1 = \frac{\frac{1}{2}V_{GB}}{V_C + \frac{1}{2}V_{GB}}, \quad \kappa_2 = \frac{\frac{1}{2}V_P}{V_C + \frac{1}{2}V_{GB}}. \tag{41}$$

Here, the factor ½ is due to the natural suggestion that the porous and second-phase crystallite parts compose the joint GB part between neighboring first-phase crystallites parts. In case of $d_{C2} \ll d$ and $d_P \ll d$ the weights $\kappa_1, \kappa_2$ equal to $(\kappa_1, \kappa_2)|_{d_{C2}, d_P \ll d} = \frac{1}{2}(V_{GB}, V_P)/V_C$, which can be roughly presented in the form suggested for the first time in [39, 40], $(\kappa_1, \kappa_2)|_{d_{C2}, d_P \ll d} = (n, m)\frac{b}{d}$  $(n, m) \sim (d_{C2} + d_P, d_P)/b$, with a some constants $n, m$, which take into account the average distance between grains and highly depends on a preparation of the GB states. The part of the volume of GB grows with a decrease in the crystallite diameter $d$, which means an increasing in the softening factor. The modified model, which takes GB into account, including theirs pores structure, leads to the following dependence of integral flow stress:

$$\sigma_\Sigma(\varepsilon) = (1 - \kappa_1)\sigma_C(\varepsilon, d) + (\kappa_1 - \kappa_2)\sigma_{C2}(\varepsilon, d_{C2}) - \kappa_2 \sigma_P(\varepsilon, d_P)$$
$$= \frac{V_C}{V_C + \frac{1}{2}V_{GB}}\sigma_C(\varepsilon, d) + \frac{\frac{1}{2}V_{C2}}{V_C + \frac{1}{2}V_{GB}}\sigma_{C2}(\varepsilon, d_{C2}) - \frac{\frac{1}{2}V_P}{V_C + \frac{1}{2}V_{GB}}\sigma_P(\varepsilon, d_P). \tag{42}$$

Here, $\sigma_C(\varepsilon, d)$ is the stress for the first phase – being the basic grains from the sample with diameter $d$; $\sigma_{C2}(\varepsilon, d_{C2})$ and $\sigma_P(\varepsilon, d_P)$ correspond to the stresses for crystallites and pores from the GB region of the respective average sizes $d_{C2}$ and $d_P$. The weights of the phases have to be such that $\sigma_\Sigma(\varepsilon) > 0$ for all $0 < \varepsilon < \bar{\varepsilon}$, for which $\sigma_\Sigma(\bar{\varepsilon}) = 0$. For $\kappa_1 = \kappa_2$ all the area of the second phase is filled by pores of different diameters. For $\kappa_2=0$ the second phase represents a set of the crystallites of different sizes with $b < d_{C2} \ll d$. The last case appears by the model one. The numeric relation among the average volumes $V_{C2}, V_P$ and diameters $d_{C2}, d_P$ can be established in case of the uniform distribution both for the crystallites and for pores from the second phase around the basic crystallite. Thus, for $V_{C2} = 0$, from $V_P = N_P V_P^e$, where $V_P^e \approx \frac{\pi}{6}d_P^3$ is the volume of one average pore and $N_P$ is the number of such pores on the surface of the basic crystallite (approximately, to be 2D sphere), with area $S_C = \pi d^2$. Therefore, we have

$$N_P = S_C/K_P S_P = \pi d^2 / (K_P * \frac{\pi}{4}d_P^2) = \frac{4}{K_P}\left(\frac{d}{d_P}\right)^2 \Rightarrow d_P = K_P \frac{3}{2\pi}\frac{V_P}{d^2}, \quad K_P \geq 1, \qquad (43)$$

where $S_P$ is the area of the 2D circle of the radius $d_P/2$. For $K_P = 1$ the pores completely cover the surface of the basic crystallite with minimal diameter $d_{Pmin} = \frac{3}{8}\frac{V_P}{d^2}$. For $V_P = 0$, we have a similar estimate for the size $d_{C2}$ of second-phase crystallites: $d_{C2} = K_{C2}\frac{3}{8\pi}\frac{V_{C2}}{d^2}$, $K_{C2} \geq 1$. In general, the diameters $d_{C2}, d_P$ are estimated from the system of cubic equations:

$$(V_P, V_{C2}) = N_{(P,C2)}(V_P^e, V_{C2}^e) = \frac{S_C*(V_P^e,V_{C2}^e)}{(K_P S_P + K_{C2} S_{C2})} = \frac{2\pi d^2 *(d_P^3, d_{C2}^3)}{3(K_P d_P^2 + K_{C2} d_{C2}^2)}, \quad K_P, K_{C2} \geq 1. \qquad (44)$$

In the case of absence of second-phase crystallites ($\kappa_1 = \kappa_2$) under a uniform distribution of the pores (for the composite grain) with respect to the size in the range $[V_{Pmin}, V_{Pmax}]$, which approximately corresponds to the size $d_{Pmin}, d_{Pmax}$ determined according to (43). e.g. for $K_P = 1$, in terms of the module of Burgers vector $b$ as $[d_{Pmin}, d_{Pmax}] = [n_{min}, n_{max}] * b$ with integers $[n_{min}, n_{max}] = \frac{3}{2\pi d^2}[V_{Pmin}, V_{Pmax}]$ the FS of homogeneous two-phase PC material follows from the equations (20) and (42):

$$\sigma_\Sigma(\varepsilon) = \sigma_0(\varepsilon) + \frac{V_C}{V_C + \frac{1}{2}\sum_{i=n_{min}}^{i=n_{max}} V_P^i}\left\{\alpha m \frac{Gb}{d}\sqrt{\frac{6\sqrt{2}}{\pi}m_0 \varepsilon M(0)}\left(e^{M(\varepsilon)b/d} - 1\right)^{-\frac{1}{2}}\right\} - \sigma_{\Sigma P}(\varepsilon, d_P)$$

$$\sigma_{\Sigma P}(\varepsilon, d_P) = \sum_{i=n_{min}}^{n_{max}} \frac{\frac{1}{2}V_P^i}{V_C + \frac{1}{2}V_P^i}\sigma_P^{(i)}(\varepsilon, d_P^i), \quad \sigma_P^{(i)}(\varepsilon, d_P^i) = \left\{\alpha m \frac{G}{i}\sqrt{\frac{6\sqrt{2}}{\pi}m_0 \varepsilon M_P(0)}\left(e^{M_P(\varepsilon)/i} - 1\right)^{-\frac{1}{2}}\right\} \qquad (45)$$

where a constant $\sigma_0(\varepsilon)$ is extracted from the first and second phases of the PC sample. Note, that the ratio $V_C/V_P = dK_P/4d_P$, which determines the weights of phases for single-mode crystallites and pores, is equal to 1, in case $d_P = \frac{1}{4}K_P d$ (in particular, $d_P = d$ for $V_C/V_P = 1$, when the pores compose one layer around any of the first phase crystallite filled on 1 quarter: $K_P = 4$). For α-Fe, when $n=60$ (that corresponds to smaller-angle grain boundaries in the coarse-grain limit, but not in the NC region) the stress-strain curves for large $d$ (the plots 1 and 2 on the Fig. 7) is non-significantly decreased with respect to the stress (quantitatively corresponding to the results of the experiments on Armco-iron [49]), whereas in the region of PC materials, starting from UFG materials, a negative input from the second GB phase becomes co-measurable with the input from the first phase (for basic crystallites), decreasing the values of stress and narrowing the zone of plasticity – with the range of strain ε under PD down to fracture. In particular, for the average size of pores $d_P = 60\,b = 14.9$ nm, taking from (45) the second term for $\sigma_{\Sigma P}(\varepsilon, d_P)$ with $i=60$, ($n_{max} = n_{min}$) one obtains the weights of phases calculated from (41), (43) for the maximal first phase grain $d_m = 40{,}6$ nm ($V_C = \frac{\pi}{6}d^3 = 3.5 *10^{-23} m^3$, $V_P = 5.13 * 10^{-23} m^3$): $(1 - \kappa_1) = 0.577$, $\kappa_1 = 0.423$ the formal maximal value $\sigma_{\Sigma max}(d_m, 300) = (0.577*13.27-0.423*8{,}21) = 4.18$ GPa decreases in 3,2 times with respect to $\sigma_{max}(d_m, 300)=13{,}27$ GPa, whereas the actual (in the 2-phase model) maximum $\bar{\sigma}_{\Sigma max}(d_m, 300)$ is already reached for $\bar{\varepsilon}_m > \varepsilon_m$, which however cannot be realized due to the Backofen–Considére condition with $\varepsilon_{\Sigma fr\,cond}(d_m, 300) > \varepsilon_{fr\,cond}(d_m, 300) \approx 0.135$. The reachable absolute (on $d, \varepsilon$) maximum (conditional ultimate stress) calculated at $\varepsilon_{fr\,cond} \approx 0.135$ are evaluated as

$$\sigma_S = \sigma_\Sigma(d_m, 300, \varepsilon_{fr\,cond}) = (0.577*12.79-0.423*9{,}48) = 3.37 \text{ GPa.}$$

The yield strength and the weight of the second phase for the above-mentioned parameters of the materials are equal $\sigma_{\Sigma y|d_m} = \sigma_\Sigma(0{,}002) = 0{,}335$ GPa and 42.3 % respectively. Let's collect the above calculated results for the two-phase model for α-Fe, in the table form:

| α-Fe | $d_P$, nm | $d = d_m$, nm | $V_P$, $10^4$ nm$^3$ | $V_C$, $10^4$ nm$^3$ | $\kappa_1$ | $(1-\kappa_1)$ | $\sigma_{\Sigma max}$, GPa | $\sigma_{\Sigma yd_m}$, GPa | $\sigma_S$, GPa | $\varepsilon_{\Sigma fr\,cond}$ |
|---|---|---|---|---|---|---|---|---|---|---|
|  | 14.9 | 40.6 | 5.13 | 3.5 | 42.3 % | 57.7 % | 4.18 | 0.335 | 3.37 | <0.165 |

**Table 5**: The values of average diameters $d_P$, $d$, volumes, weights formal and real ultimate stress and yield strength for the two-phase single-mode PC model of a α-Fe at $T=300K$ for extremal (with respect to ε) grain with nanopores.

It should be noted that the model has to be modified in a significant way to describe stress-strain dependence "$\sigma_\Sigma - \varepsilon$" due to excitations of new dislocation ensembles with other Burgers vectors in new zones of localized plasticity with a growth of deformation, where other material of a material will originate.

Now we turn to the influence (more investigated theoretically) of the second GB phase on the plastic and strengthened properties of PC materials.

### 10. Generalized Hall-Petch law for two-phase single-mode α-Fe, Cu, Al, Ni, α-Ti, Zr

Note, in the first place, that the flow stress[10] $\sigma_\Sigma(\varepsilon)$ in (42), (20) can be naturally enlarged in the case of dispersion hardening,

$$\sigma_{\Sigma dis}(\varepsilon) = (1 - U_{dis})\{f_1\sigma_C(\varepsilon_C) + f_2\sigma_{C2}(\varepsilon_{C2}) - f_3\sigma_P(\varepsilon_P)\} + U_{dis}\sigma_{dis}(\varepsilon_{dis}, d_{dis})$$
$$\varepsilon = (1 - U_{dis})(f_1\varepsilon_C + f_2\varepsilon_{C2} + f_3\varepsilon_P) + U_{dis}\varepsilon_{dis}, \qquad 0 \leq U_{dis} << 1, \qquad \sum_{i=1}^{3} f_i = 1 \quad (46)$$

by a "third phase" term: $\sigma_{dis}(\varepsilon, d_{dis})$ with the weight $U_{dis}$ for $d_{dis}$ being by average linear size of particles (from other compounds), which realize this hardening process. Here, the role of $\sigma_{dis}$ is analogous to that of $\sigma_{C2}(\varepsilon, d_{C2})$ for the second phase grains, but with a proper shear modulus $G_{dis}$ and Burgers vector $b_{dis}$ now distributed inside the first phase grains. For CG and other PC samples with $d_{dis} \ll d$ the particles can provide a growth of integral FS: $\sigma_{\Sigma dis}(\varepsilon)$, in particular, yield strength, when $Gb^3 < G_{dis}b_{dis}^3$. But in NC samples hardening can also take place under more complicated conditions, $d_0 < d_{dis} < d$, and possibly $Gb^3 > G_{dis}b_{dis}^3$, while un-hardening is possible for $d_{dis} < d_0 < d$ with a certain relation among unit dislocation energies $\frac{1}{2}Gb^3$, $\frac{1}{2}G_{dis}b_{dis}^3$ following the above results, see e.g. [38, 39, 40]. In changing the sizes of $d, d_{C2}, d_P, d_{dis}$ the correspondence among $\varepsilon_C, \varepsilon_P, \varepsilon_{C2}, \varepsilon_{dis}$ change as well, so that in order to theoretically determine, e,g, $\varepsilon = 0.002$ in a three-phase model for every distribution of $d, d_C, d_P, d_{dis}$ one should use the one-phase model for each phases. In [41], we began to study dispersion hardening (by Cu-particles) and intend to continue this research.

To determine the values of the constant $m_0$ (24) for the two-phase model let us use the known experimental values for HP coefficient $k(0,002)$ in single-mode PC samples with BCC, FCC and HCP CL from Table 1 with small-angle GBs, corresponding to the values of $\sigma_0$, $G$, lattice constants $a$ [45], Burgers vectors with the least possible lengths $b$, with respective most realizable sliding systems (given by Table 2), interaction constant for dislocation $\alpha$ [5, 23] and the computed values of the extreme (crystalline, i.e. first phase) $d_0$ and new (two-phase) $d_{\Sigma 0}$ grain sizes, maximal differences of yield strength $\Delta\sigma_m$, $\Delta\sigma_{\Sigma m}$ in accordance with (22), (23) and (41), (42) for $T=300K$:

| Type of CL | BCC | FCC | | | HCP | |
|---|---|---|---|---|---|---|
| Material | α-Fe | Cu | Al | Ni | α-Ti | Zr |
| $d_0$, nm | 23.6 | 14.4 | 13.6 | 22.6 | 23.8 | 28.0 |
| $\Delta\sigma_m$, GPa | 2.29-2.69 | 1.31 | 0.83 | 1.18 | 1.58-1.79 | 0.99 |
| $d_{\Sigma 0}$, $d_{Ps}$ | 23.5 | 14.3 | 13.5 | 22.5 | 23.5 | 27.8 |

---

[10] We restrict ourselves [39,40] in the two-phase model by the case when the strain values corresponding to the first-phase crystallites $\varepsilon_C$, pores $\varepsilon_P$ and second-phase grains $\varepsilon_{C2}$ coincide with each other, i.e., all the phases of the PC materials are transformed in a coherent manner as non- interacting objects. In general, the strain of each phase is heterogeneous, so that in addition to the integral (flow) stress $\sigma_\Sigma(\varepsilon)$ (41), (42): $\sigma_\Sigma(\varepsilon) = f_1\sigma_C(\varepsilon_C) + f_2\sigma_{C2}(\varepsilon_{C2}) - f_3\sigma_P(\varepsilon_P)$ for the volume parts $(f_1, f_2, f_3) = \left(\frac{V_C}{V_C + \frac{1}{2}V_{GB}}, \frac{\frac{1}{2}V_{C2}}{V_C + \frac{1}{2}V_{GB}}, \frac{\frac{1}{2}V_P}{V_C + \frac{1}{2}V_{GB}}\right)$ according to the model of equal stresses the integral strain ε: one should be determined in the same way, as $\varepsilon = f_1\varepsilon_C + f_2\varepsilon_{C2} + f_3\varepsilon_P$, according to the model of equal strains, when all compounds reach the own yield strength as for the case of steel with ferrite-perlite structure [43,44].

| *nm* | $d_{Pg}$ | 22.5 | | 13.5 | | 11.4 | | 18.9 | | 22.5 | | 25.0 | |
|---|---|---|---|---|---|---|---|---|---|---|---|---|---|
| | $\bar{d}_P$ | ~400 | 24.0 | ~170 | 14.4 | ~160 | 13.6 | ~400 | 22 | ~400 | 24.8 | ~400 | 28 |
| $\Delta\sigma_{\Sigma m}$, GPa | $-\sigma_{Pms}$ | 2.21 | 0.09 | 1.29 | 0.05 | 0.80 | 0.03 | 1.16 | 0.04 | 1.51 | 0.06 | 0.99 | 0.04 |
| | $-\sigma_{Pmg}$ | 1.91 | 0.45 | 1.11 | 0.22 | 0.69 | 0.13 | 0.99 | 0.20 | 1.31 | 0.26 | 0.83 | 0.19 |
| | $-\bar{\sigma}_{Pm}$ | 0.52 | 0.98 | 0.31 | 0.65 | 0.19 | 0.39 | 0.27 | 0.60 | 0.35 | 0.83 | 0.23 | 0.47 |

**Table 6**: The values $\Delta\sigma_m = (\sigma_m - \sigma_0)$, $\Delta\sigma_{\Sigma m} = (\sigma_{\Sigma m} - \sigma_0)$, and negative inputs in $\sigma_\Sigma$ from different pores: $-\sigma_{Pms}$, $-\sigma_{Pmg}$, $-\bar{\sigma}_{Pm}$ in integral yield strengths for BCC, FCC and HCP PC metal samples with $d_0$, $b$, $G$, taken from Table 1 at $\varepsilon = 0.002$ and $d_{\Sigma 0}$ obtained from Fig. 8 for $(d_{Ps}, d_{Pg}) = (0.02, 0.10)*d$, which correspond to average weights of the phases $(f_1, f_3)$; $(1-\kappa_1, \kappa_1)$; equal to (0.962; 0.038) for small- and (0.833; 0.167)[11] for large-angle and constant size of porous $\bar{d}_P$ for each PC samples, in case it does exist. The lowest boundary $d_{LB}$ for the existence of two-phases single-mode PC samples with $\bar{d}_P$ are estimated as $d_{LB}(\alpha\text{-Fe; Cu; Al; Ni; }\alpha\text{-Ti; Zr}) \approx (50;29;30;42;44;50)$ nm, with the admissible weights of porous part. $f_3(\alpha\text{-Fe; Cu; Al; Ni; }\alpha\text{-Ti; Zr}) \approx (0.49; 0,50; 0.48; 0.51; 0.35; 0.53)$.

We accept the same values for the polyhedral parameter $m_0$, and therefore for HP coefficients $k(0.002)$ as ones for pure crystalline phase from the Table 1. The theoretical Hall-Petch curves for integral yield strengths are presented in the Figure 8.

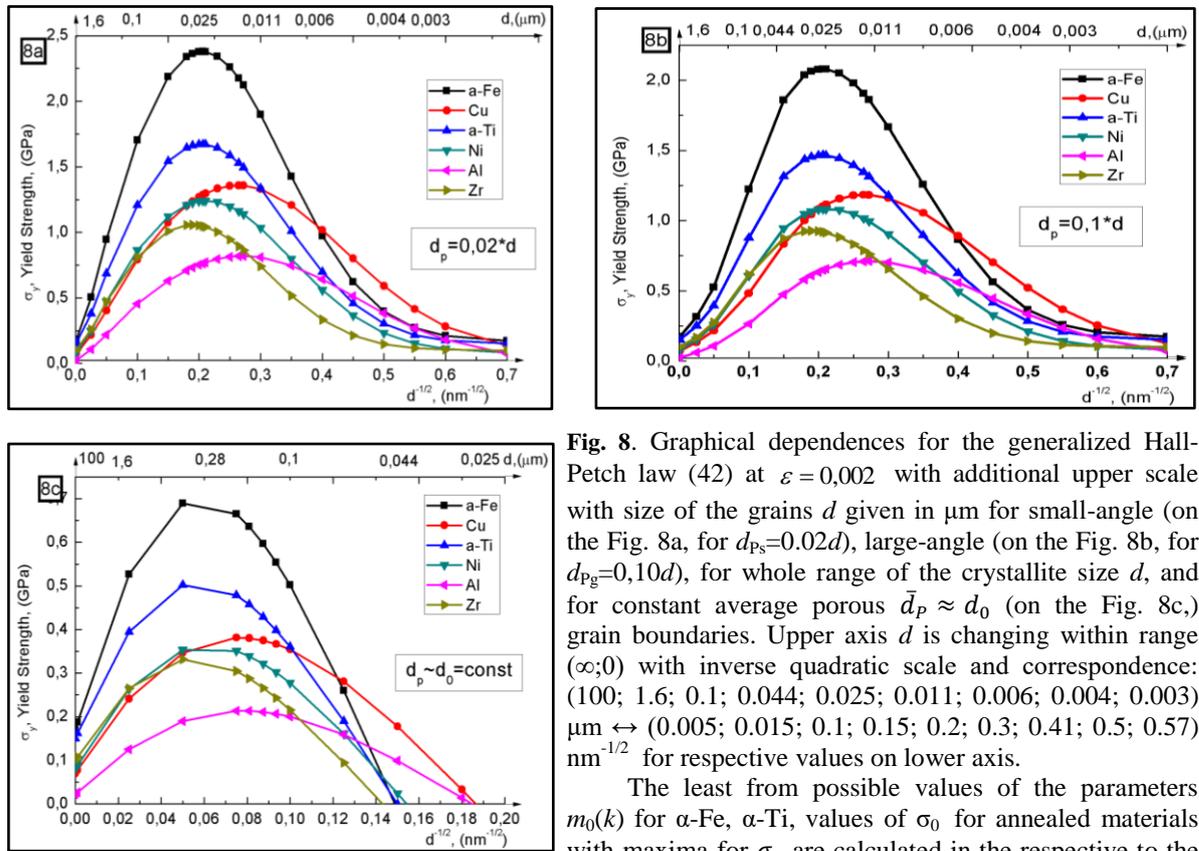

**Fig. 8**. Graphical dependences for the generalized Hall-Petch law (42) at $\varepsilon = 0.002$ with additional upper scale with size of the grains $d$ given in μm for small-angle (on the Fig. 8a, for $d_{Ps}=0.02d$), large-angle (on the Fig. 8b, for $d_{Pg}=0,10d$), for whole range of the crystallite size $d$, and for constant average porous $\bar{d}_P \approx d_0$ (on the Fig. 8c,) grain boundaries. Upper axis $d$ is changing within range $(\infty;0)$ with inverse quadratic scale and correspondence: (100; 1.6; 0.1; 0.044; 0.025; 0.011; 0.006; 0.004; 0.003) μm ↔ (0.005; 0.015; 0.1; 0.15; 0.2; 0.3; 0.41; 0.5; 0.57) nm$^{-1/2}$ for respective values on lower axis.

The least from possible values of the parameters $m_0(k)$ for α-Fe, α-Ti, values of $\sigma_0$ for annealed materials with maxima for $\sigma_y$ are calculated in the respective to the Table 1 extreme grain sizes $d_0$

From the plots in the Fig. 8 it follows, that with growth of the second-phase parts the extreme size of the grains: $d_{Ps}$, $d_{Pg}$ and maxima for yield strengths $\sigma_{\Sigma m}$ are decreased both for the small- and large-angle GB, whereas for non-constant values of the second phase input with diameter of pores $\bar{d}_P \sim d_0$ the extreme values $\boldsymbol{d_{\Sigma 0}}$ is shifted into regions with large grains, $d_{\Sigma 0} \sim 5* d_0$ with decreasing of the value $\boldsymbol{\sigma_{\Sigma m}}$ for all two-phase single-mode PC samples. For the materials with constant size of porous $\bar{d}_P$ the lowest boundaries $d_{LB}$ for the existence of the two-phases single-mode PC samples arise.

---

[11] In calculating the weights of hard and weak phases we have used the relations $(1-\kappa_1) \equiv f_1 = V_C/(V_C + \frac{1}{2}V_P)$ for $V_{GB} = V_P$ according to (41)-(44), whereas in [39] a more rough approximation have been used related with $(f_1, f_3) = \left(1 - \frac{nb}{d}, \frac{nb}{d}\right) = (d^3/(d+d_P)^3, \{(d+d_P)^3 - d^3\}/(d+d_P)^3)$. .

## 11. Temperature dependence of yield strength and extreme grain sizes for two-phase Al

Here, we continue the study of the temperature-dimension effect revealed within one-phase PC model for Al in Sec. 7 within the same realizations for the two-phase model as in the previous Section, i.e. for small-angle, large-angle GB and with constant size of pores in the whole range of average diameter of the first-phase grains $d$, without second-phase grains (for simplicity). Based on the $T$-dependence study of pure crystalline phase of the single-mode PC material, described by the Eqs. (32), we add to $\sigma_C(T)$ according to (42), (45) the negative input of a similar to (32) $T$-dependence of the second-phase, initiated by the porous part (with except for $\sigma_0(T)$, which was already included in $\sigma_C(T)$). The results of the theoretical study of the $T$-dependence for the Hall-Petch law for Al samples for small-, large-angle GB and for the constant size of porous from the second phase in the whole range of diameters $d$ of the first phase crystallites in comparison with pure crystallite phase realization [39] are given by Table 7 and Fig. 9.

| Al $T,K$ | $d_0$, nm | $\Delta\sigma_m$, GPa | $d_{\Sigma 0}$, nm; $(d_{Ps}=0,02d)$ | $\Delta\sigma_{\Sigma m}$, GPa | $d_{\Sigma 0}$, nm; $(d_{Pg}=0,10d)$ | $\Delta\sigma_{\Sigma m}$, GPa | $d_{\Sigma 0}$, nm; $(\bar{d}_P=13,6nm)$ | $\Delta\sigma_{\Sigma m}$, GPa |
|---|---|---|---|---|---|---|---|---|
| 350 | 11.3 | 0.85 | 11.1 | 0.84 | 10.5 | 0.73 | 175.0 | 0.18 |
| 300 | 13.6 | 0.83 | 13.0 | 0.80 | 12.0 | 0.69 | 160.0 | 0.19 |
| 250 | 16.8 | 0.74 | 16,0 | 0.73 | 15.5 | 0.64 | 150.0 | 0.21 |
| 200 | 21.5 | 0.67 | 20,3 | 0.65 | 19.5 | 0.58 | 125.0 | 0.23 |
| 150 | 29.5 | 0.59 | 28,0 | 0.57 | 27.5 | 0.51 | 125.0 | 0.26 |

**Table 7** The values $d_0$, $\Delta\sigma_m$ for PC Al samples with $d_0, b, G$ for $T=300K$ are taken from Table 1, Fig. 5 at ε=0,002 and $d_{\Sigma 0}, \Delta\sigma_{\Sigma m}=(\sigma_{\Sigma m}-\sigma_0)$ obtained from the Fig. 9 for $d_{Ps}=0,02*d$ and $d_{Pg}=0,10*d$, that corresponds to average weights of the phases $(\bar{f}_1, \bar{f}_3)$ = (0.962; 0.038) and (0.833; 0.167) without account for the weak phase grains for respective small- and large-angle GBs and constant size of porous $\bar{d}_P$ PC Al samples, when it exists. The boundary of existence $d_{LB}$ is estimated from 22.5 nm to 39 nm in the range [150K, 350K].

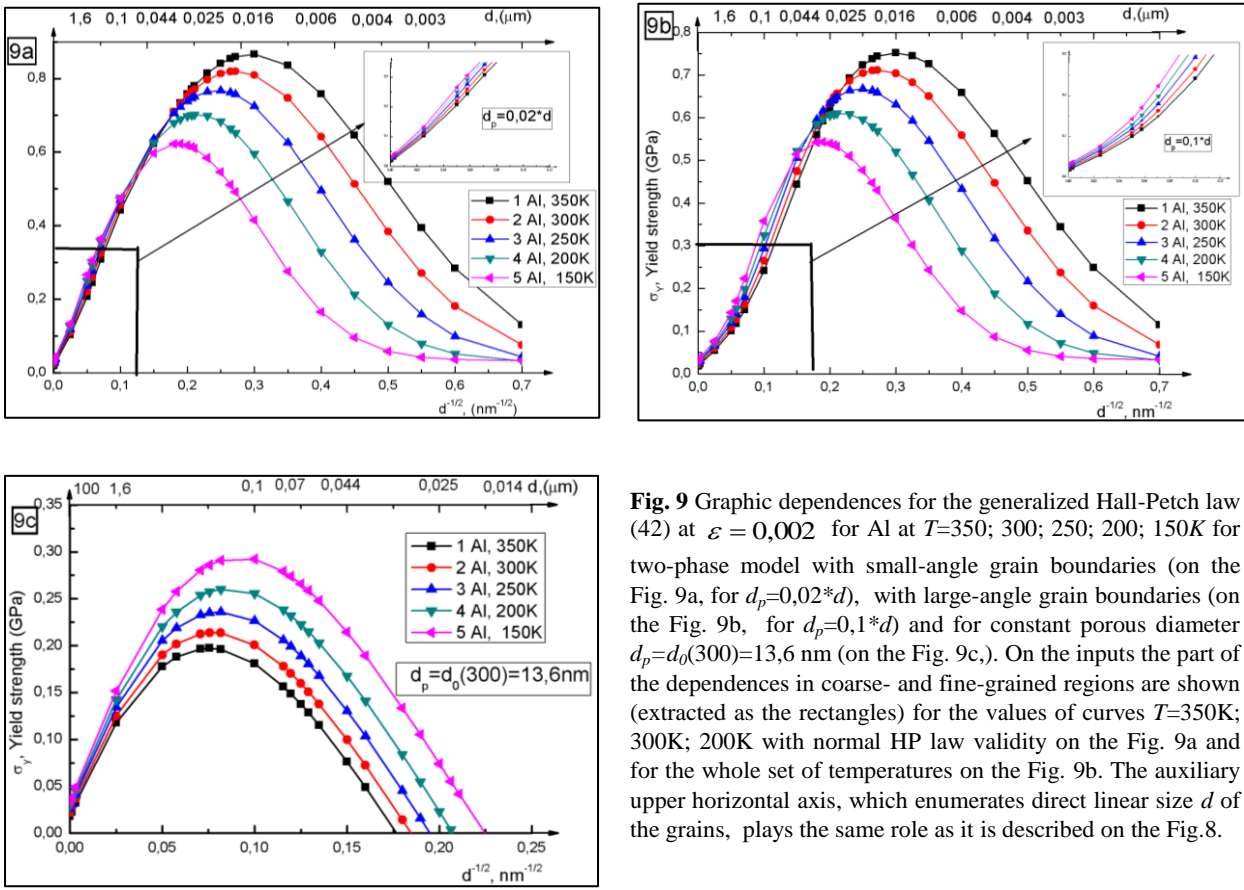

**Fig. 9** Graphic dependences for the generalized Hall-Petch law (42) at $\varepsilon = 0,002$ for Al at $T=350$; 300; 250; 200; 150$K$ for two-phase model with small-angle grain boundaries (on the Fig. 9a, for $d_p=0,02*d$), with large-angle grain boundaries (on the Fig. 9b, for $d_p=0,1*d$) and for constant porous diameter $d_p=d_0(300)=13,6$ nm (on the Fig. 9c,). On the inputs the part of the dependences in coarse- and fine-grained regions are shown (extracted as the rectangles) for the values of curves $T$=350K; 300K; 200K with normal HP law validity on the Fig. 9a and for the whole set of temperatures on the Fig. 9b. The auxiliary upper horizontal axis, which enumerates direct linear size $d$ of the grains, plays the same role as it is described on the Fig.8.

From the analysis of the Figure 9, it follows that the TDE for two-phase PC single-mode Al aggregates both for small-angle (Fig. 9a) and for large-angle (Fig. 9b) GB till takes place, but with decreasing of the critical value size for the grains, being estimated as $d_{\Sigma 1}<3d_{\Sigma 0}$ due to unhardening from the second phase.

For the case of constant pores (Fig. 9c) with $d_P = d_0(300)$ the effect is completely neutralized and the PC material has the standard T-behaviour as for CG and till to SMC (and NC if it exists) aggregates. The reason of the above neutralization is due to shifting from the extremal NC region for the subcritical grain for composite model of grains so that the new extremal values $d_{\Sigma 0}$ lie in the range 125-175 nm being more than critical for TDE diameter $d_1 \sim 40$ nm. The first part of TDE disappears as well, because of the decreasing of the extremal grain size $d_{\Sigma 0}$ with decreasing of the temperature[12].

### 12. Stress-strain curves for the two-phase two-mode α-Fe with different grain boundaries

Once again, as in Section 8, the dependence (42) in the form (45) but with explicit second-mode crystallite part (extracted, e.g. from the non-porous part of grain boundary region) of $\sigma_\Sigma(\varepsilon)$, together with the stress-strain curve plot $\sigma_\Sigma = \sigma_\Sigma(\varepsilon)$ in Figs. 10, 11, permits us to find the strain hardening coefficient $\theta_\Sigma(\varepsilon)$, assuming that $\theta_{0\Sigma}(\varepsilon) = d\sigma_0/d\varepsilon$,:

$$\sigma_\Sigma(\varepsilon) = \sum_{i=1}^2 f_i \sigma_{Ci}(\varepsilon_i, d_i, T) - \sum_{i=1}^2 f_{Pi}\sigma_{Pi}(\varepsilon_{Pi}, d_{Pi}, T), \quad \varepsilon = \sum_i (f_i \varepsilon_i + f_{Pi}\varepsilon_{Pi}), \quad \sum_i (f_i + f_{Pi}) = 1, \quad (47)$$

$$\theta_\Sigma(\varepsilon) = \sum_{i=1}^2 f_i \theta_{Ci}(\varepsilon_i) - \sum_{i=1}^2 f_{Pi}\theta_{Pi}(\varepsilon_{Pi}), \quad \theta_{Ci}(\varepsilon_i) = \frac{d\sigma_{Ci}(\varepsilon_i)}{d\varepsilon_i}, \quad \theta_{Pi}(\varepsilon_{Pi}) = \frac{d\sigma_{Pi}(\varepsilon_{Pi})}{d\varepsilon_{Pi}}, \quad (48)$$

for $\theta_{Ci}(\varepsilon_i), \theta_{Pi}(\varepsilon_{Pi})$ being by the strain hardening coefficients determined according to (36) respectively for the basic crystallite of diameter $d_1$, for the second-mode crystallites of diameter $d_2$ and for the porous parts of diameters $d_{P1}$, $d_{P2}$ for respective grains under assumption of homogeneous character of strain for both phases, i.e. for $\varepsilon = \varepsilon_i = \varepsilon_{Pi}$, as it was supposed in the footnote 10. The second-mode grains have the structure being analogous to the first-mode ones, according to the Fig. 3b. For simplicity we consider both of the crystallites without the second-phase GB grains, but with the same diameters of pores, $d_{P1} = d_{P2}$, for both grain's modes. The stress-strain curves for the dependences $\sigma_\Sigma(\varepsilon) = \sigma_\Sigma(\varepsilon, d_1, d_2, T)$ (47) for a two-phase PC α-Fe sample at $T$=300K for various two-mode average grain sizes: $(d_1, d_2) = (d, 0.5 * d)$, with weights for composite first-mode and second-mode grains: $f_1 + f_{P1}$=70% and $f_2 + f_{P2} = 30\%$, for small-, large-angle GB from the 2-nd phase and for $\sigma_\Sigma(\varepsilon) = \sigma_\Sigma(\varepsilon, d_1, 0, T)$ for only single-mode grains are given by Fig. 10, 11 on the basis of Table 8. The values of conditional elastic limit $\sigma(0,0005)$ are formally calculated according (47) without account for $\sigma_0(\varepsilon)$.

| α-Fe | $(d_1, d_2) = (d, 0.5 * d), nm, \ d_{P1s} \in [0.001*d, 0.025*d], nm, \ \sigma_\Sigma(\varepsilon)$, GPa | | | | | | | | | |
|---|---|---|---|---|---|---|---|---|---|---|
| $\varepsilon, \times 10^{-2}$ <br> $d; d_{P1}$ | 0,05 | 0,1 | 0,2 | 0,5 | 1 | 2 | 5 | 10 | 20 | 30 |
| $10^6; 10^3$ | 0,009 | 0,182 | 0,188 | 0,198 | 0,210 | 0,225 | 0,254 | 0,281 | 0,308 | 0,320 |
| $10^6; d_2 = 0$ | 0.008 | 0.181 | 0.186 | 0.195 | 0.206 | 0.220 | 0.246 | 0.270 | 0.294 | 0.305 |
| $10^5; 2*10^2$ | 0,03 | 0,21 | 0,23 | 0,26 | 0,29 | 0.34 | 0.43 | 0,51 | 0,59 | 0,63 |
| $10^5; d_2 = 0$ | 0.02 | 0.20 | 0.22 | 0.25 | 0.28 | 0.32 | 0.40 | 0.48 | 0.55 | 0.59 |
| $10^3; 20$ | 0.23 | 0.39 | 0.63 | 0,90 | 1,20 | 1,61 | 2.38 | 3,15 | 4,01 | 4.47 |
| $10^3; d_2 = 0$ | 0.21 | 0.47 | 0.60 | 0.85 | 1.13 | 1.52 | 2.23 | 2.94 | 3.72 | 4.13 |
| 150; 3.0 | 0,66 | 1,14 | 1,55 | 2,34 | 3,21 | 4,40 | 6,53 | 8,43 | 10.06 | 10.45 |
| 150; $d_2 = 0$ | 0.61 | 1.07 | 1.44 | 2.17 | 2.98 | 4.09 | 6.07 | 7.87 | 9.48 | 9.97 |
| 23.6; 0.5 | 1,02 | 1,62 | 2,22 | 3,38 | 4,63 | 6,26 | 8,80 | 10,31 | 9,77 | 7,72 |
| 23.6; $d_2 = 0$ | 1.09 | 1.72 | 2.36 | 3.60 | 4.95 | 6.73 | 9.63 | 11.57 | 11.57 | 9.62 |
| 10; 0.25 | 0.61 | 1,03 | 1,39 | 2,06 | 2,77 | 3,62 | 4.64 | 4.61 | 3,05 | 1,60 |
| 10; $d_2 = 0$ | 0.78 | 1.27 | 1.72 | 2.59 | 3.50 | 4.61 | 6.01 | 6.12 | 4.16 | 2.19 |
| $d_{P1g} \in [0.017*d, 0.25*d], nm, \sigma_\Sigma(\varepsilon)$, GPa | | | | | | | | | | |
| $10^6; 1.7*10^4$ | 0.000 | 0.172 | 0.176 | 0.183 | 0.192 | 0.203 | 0.224 | 0.243 | 0.263 | 0.271 |
| $10^6; d_2 = 0$ | 0.001 | 0.173 | 0.177 | 0.184 | 0.192 | 0.203 | 0.222 | 0.241 | 0.259 | 0.268 |
| $10^5; 4*10^3$ | 0.002 | 0.176 | 0.184 | 0.202 | 0.221 | 0.247 | 0.295 | 0.340 | 0.385 | 0.405 |

---

[12] A result similar to the one given by Fig. 9c, but for single-mode PC samples of Cu within molecular dynamic simulations was obtained in [22] (see Fig. 5 therein), but in the NC range, instead of the SMC range and for non-quasistatic PD with $\dot\varepsilon = 5 * 10^8 s^{-1}$..

| | | | | | | | | | |
|---|---|---|---|---|---|---|---|---|---|
| $10^{\backslash 5}$; $d_2 = 0$ | 0.000 | 0.179 | 0.188 | 0.206 | 0.225 | 0.252 | 0.300 | 0.346 | 0.391 | 0.411 |
| $10^3$; $10^2$ | 0.004 | 0.25 | 0.30 | 0.39 | 0.49 | 0.64 | 0.91 | 1.18 | 1.51 | 1.72 |
| $10^3$; $d_2 = 0$ | 0.067 | 0.28 | 0.33 | 0.44 | 0.57 | 0.73 | 1.05 | 1.36 | 1.79 | 1.93 |
| 150; 30 | 0.01 | 0.26 | 0.33 | 0.45 | 0.59 | 0.80 | 1.24 | 1.789 | 2.70 | 3.46 |
| 150; $d_2 = 0$ | 0.11 | 0.34 | 0.43 | 0.61 | 0.81 | 1.11 | 1.69 | 2.368 | 3.37 | 4.14 |
| 23.6; 6 | 0.45 | 0.84 | 1.14 | 1.74 | 2.39 | 3.28 | 4.83 | 6.003 | 6.16 | 5.08 |
| 23.6; $d_2 = 0$ | 0.57 | 1.00 | 1.36 | 2.08 | 2.87 | 3.95 | 5.83 | 7.311 | 7.70 | 6.56 |
| 10; 2.5 | 0.35 | 0.70 | 0.94 | 1.39 | 1.87 | 2.45 | 3.17 | 3.172 | 2.11 | 1.11 |
| 10; $d_2 = 0$ | 0.49 | 0.89 | 1.20 | 1.80 | 2.43 | 3.21 | 4.19 | 4.278 | 2.91 | 1.53 |

**Table 8**: The calculated values of stress $\sigma_\Sigma(\varepsilon)$ for the two- phase of two-modal and single-modal PC aggregates α-Fe at $T$=300K for various average grain sizes within the range $\varepsilon \in [0.0005;0.3]$ respectively for small-angle (with accuracy up to the third, for $d = 10^6$ nm and to the second for $d \leq 10^5$ nm significant decimal digit), large-angle (with accuracy up to the third, for $d \geq 10^5$ nm and to the second for $d \leq 10^3$ nm significant decimal digit) GB characterized by the porous sizes $d_{P1s} = d_{P2s}$, $d_{P1g} = d_{P2g}$. to be different for CG, UFG, SMC and NC grains within the indicated ranges: [0.001*d, 0.02*d] and , [0.017*d, 0.25*d] being different for CG and SMC, NC samples.

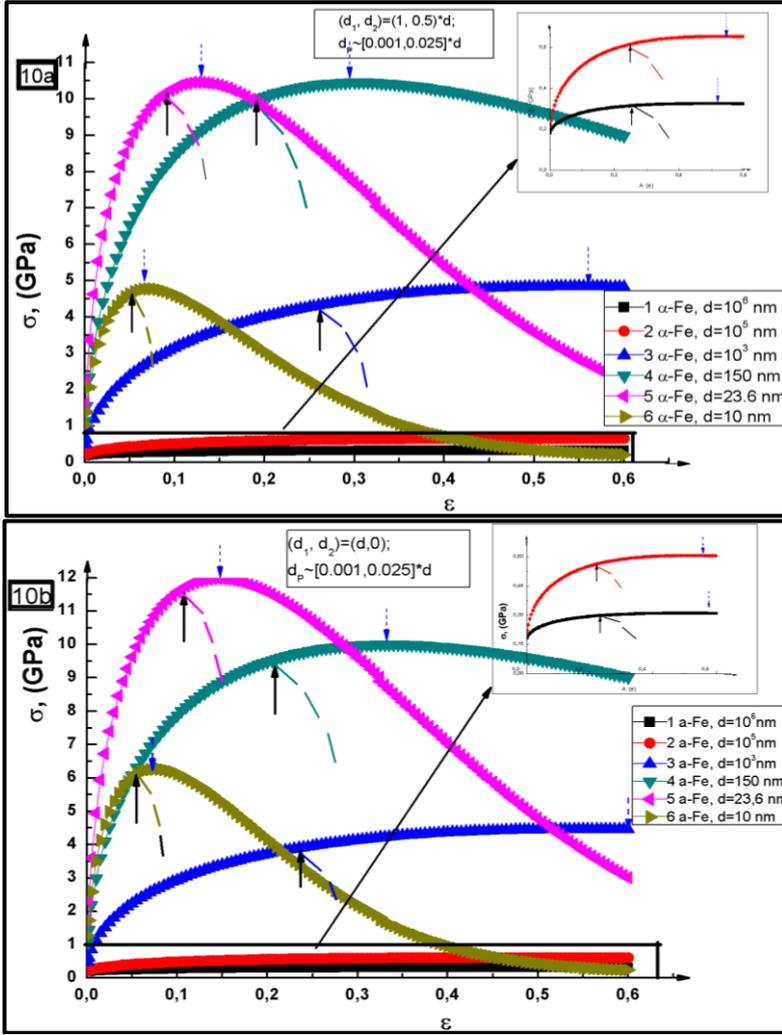

**Fig. 10.** Plotted dependence for $\sigma_\Sigma = \sigma_\Sigma(\varepsilon, d_1, d_2, T)$ in (47) for α-Fe at $T$=300K, $m_0 \cdot \alpha^2 = 3{,}66$, with stress-strain curves 1, 2, 3, 4, 5, 6 for two-modal on the Fig 10a with $(d_1, d_2) = (d, 0.5 * d)$: $d = 10^{-3}$; $10^{-4}$; $10^{-6}$m; $d$=150 nm; $d= d_0$= 23.6 nm; $d$=10 nm and single-modal on the Fig 10b for $d_2 = 0$ two-phase PC aggregates α-Fe for small-angle valued GB. By the black arrows below it is indicated the values, where the Backofen-Considére condition (49) is realized with $\varepsilon_{\Sigma fr\,cond}$ given by Table 9 and expected dashed lines of curves before the fracture. The blue arrows from top indicate the maximums $\sigma_{\Sigma max}(\varepsilon)$ of $\sigma_\Sigma(\varepsilon)$ for the strains found numerically. On the input the plots for the stress-strain curves for CG aggregates are shown. Fig. 10a and Fig. 10b are given for small-angle valued GB.

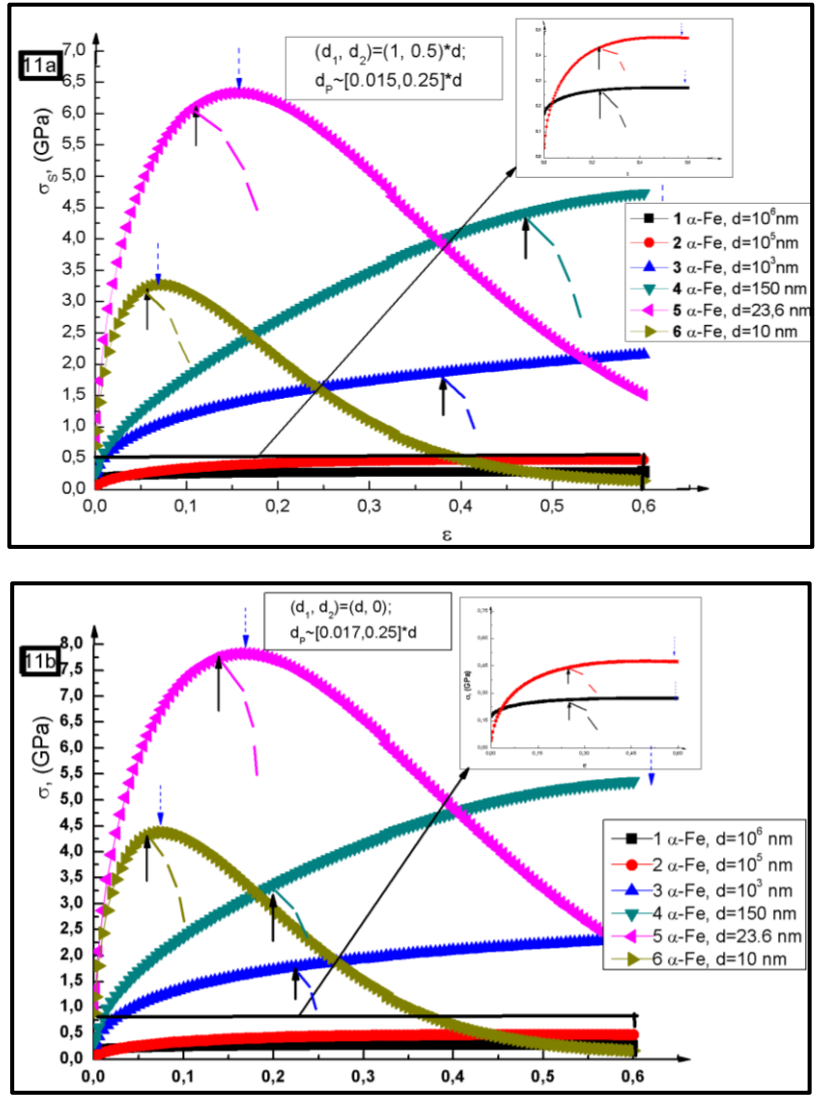

**Fig. 11**. Theoretical plots for two-phase PC aggregates α-Fe with large-angle valued GB for two-modal, $\sigma_\Sigma = \sigma_\Sigma(\varepsilon, d_1, d_2, T)$ on Fig. 11a and for single-modal samples $\sigma_\Sigma = \sigma_\Sigma(\varepsilon, d_1, 0, T)$ on Fig. 11b.

The Backofen-Considére condition now takes the form $\sigma_\Sigma = d\sigma_\Sigma/d\varepsilon = \theta_\Sigma(\varepsilon)$, which selects the regions of homogeneous and localized PD, and permits to determine the values of conditional strain $\varepsilon_{\Sigma fr\ cond}$ and stress of fracture (ultimate stress) $\sigma_{\Sigma S} = \sigma_\Sigma(\varepsilon_{\Sigma fr\ cond})$ from the equation (with allowance for $\varepsilon = \varepsilon_i = \varepsilon_{\text{P}i}$)

$$\theta_{0\Sigma}(\varepsilon) - \sigma_0(\varepsilon) = m \frac{Gb}{\sqrt{\varepsilon}} \sqrt{\frac{3}{\pi\sqrt{2}}} m_0 \alpha^2 M(0) \sum_{i=1}^{2} \left[ \frac{f_i}{d_i} \left( e^{M(\varepsilon)b/d_i} - 1 \right)^{-\frac{1}{2}} X(\varepsilon, d_i, T) - \right.$$
$$\left. - \frac{f_{Pi}}{d_{Pi}} \left( e^{M(\varepsilon)b/d_{Pi}} - 1 \right)^{-\frac{1}{2}} X(\varepsilon, d_{Pi}, T) \right], \quad (49)$$
$$X(\varepsilon, d_i, T) = 2\varepsilon - 1 + \frac{3\varepsilon}{1+\varepsilon} M(\varepsilon) \frac{b}{d_i} e^{M(\varepsilon)b/d_i} \left( e^{M(\varepsilon)b/d_i} - 1 \right)^{-1}$$

The weights $f_i$ for two-modal PC samples with small-angle GB are equal respectively for $K_{Pi} = 1, i = 1,2$ in (43), (44):

$(f_1, f_{P1}; f_2, f_{P2})_{|d=10^6 nm} = (0.7(0.998, 0.002); 0.3(0.996, 0.004)) = (0.698, 0.002; 0.299, 0.001)$
$(f_1, f_{P1}; f_2, f_{P2})_{|d=10^5 nm} = (0.7(0.996, 0.004); 0.3(0.992, 0.008)) = (0.697, 0.003; 0.298, 0.002)$ (50)
$(f_1, f_{P1}; f_2, f_{P2})_{|d \leq 10^3 nm} = (0.7(0.980, 0.020); 0.3(0.962, 0.038)) = (0.679, 0.291; 0.288, 0.012)$

and with large-angle GB in accordance with the rule (41) and footnote 11: $(f_i, f_{Pi}) = \left( \frac{V_{Ci}}{V_{Ci} + \frac{1}{2}V_{Pi}}, \frac{\frac{1}{2}V_{Pi}}{V_C + \frac{1}{2}V_{Pi}} \right)$,

$$(f_1, f_{P1}; f_2, f_{P2})|_{d=10^6 nm} = (0.7(0.967, 0.033); 0.3(0.936, 0.038)) = (0.677, 0.023; 0.281, 0.019)$$
$$(f_1, f_{P1}; f_2, f_{P2})|_{d=10^5 nm} = (0.7(0.926, 0.074); 0.3(0.862, 0.138)) = (0.648, 0.052; 0.26, 0.04)$$
$$(f_1, f_{P1}; f_2, f_{P2})|_{d=10^3 nm} = (0.7(0.833, 0.167); 0.3(0.714, 0.286)) = (0.583, 0.117; 0.214, 0.086) \quad (51)$$
$$(f_1, f_{P1}; f_2, f_{P2})|_{d=150 nm} = (0.7(0.714, 0.286); 0.3(0.555, 0.445)) = (0.500, 0.200; 0.167, 0.133)$$
$$(f_1, f_{P1}; f_2, f_{P2})|_{d\leq 23.6 nm} = (0.7(0.663, 0.337); 0.3(0.496, 0.504)) = (0.474, 0.226; 0.149, 0.151)$$

without no allowance for the weak phase grains for both modes. The weights for the single-modal PC samples $(f_1, f_{P1})$, for $(f_2, f_{P2}) = (0,0)$, are given in the brackets in (50), (51) behind the factor 0.7. The conditional strain $\varepsilon_{\Sigma fr\, cond}$ and ultimate stress $\sigma_{\Sigma S}$ for single-modal and two-modal PC samples are calculated from the equation (49) for $\theta_{0\Sigma}(\varepsilon) = \sigma_0(\varepsilon) = 0$ and presented in the Table 9

|  | $(d_1, d_2) = (d, 0.5*d)$, $d_{P1s}$ | | | | | | $(d_1, d_2) = (d, 0)$, $d_{P1s}$ | | | | | |
|---|---|---|---|---|---|---|---|---|---|---|---|---|
| $d$, nm | 10 | 23.6 | 150 | $10^3$ | $10^5$ | $10^6$ | 10 | 23.6 | 150 | $10^3$ | $10^5$ | $10^6$ |
| $\varepsilon_{\Sigma fr\, cond}$ | 0.057 | 0.095 | 0.18 | 0.265 | 0.225 | 0.225 | 0.065 | 0.115 | 0.195 | 0.225 | 0.230 | 0.230 |
| $\sigma_{\Sigma S}$, GPa | 4.73 | 10.21 | 9.87 | 4.34 | 0.606 | 0.312 | 6.22 | 11.69 | 9.43 | 3.82 | 0.562 | 0.297 |
| $\varepsilon_{\Sigma max}$ | 0.065 | 0.115 | 0.29 | 0.550 | 0,525 | 0.505 | 0.075 | 0.145 | 0.335 | 0.555 | 0.495 | 0.50 |
|  | $(d_1, d_2) = (d, 0.5*d)$, nm, $d_{P1g}$ | | | | | | $(d_1, d_2) = (d, 0)$, nm, $d_{P1g}$ | | | | | |
| $\varepsilon_{\Sigma fr\, cond}$ | 0.06 | 0.12 | 0.475 | 0.385 | 0.225 | 0.225 | 0.065 | 0.13 | 0.195 | 0.225 | 0.23 | 0.23 |
| $\sigma_{\Sigma S}$, GPa | 3.24 | 6.29 | 4.41 | 1.86 | 0.397 | 0.265 | 4.35 | 7.77 | 3.32 | 1.77 | 0.391 | 0.262 |
| $\varepsilon_{\Sigma max}$ | 0.065 | 0.143 | 0.65 | >1.0 | 0.51 | 0.505 | 0.075 | 0.145 | 0.65 | >1.0 | 0.51 | 0.50 |

**Table 9**: The calculated values from (49), (47) of ultimate stress $\sigma_{\Sigma S}$ and conditional strain fracture $\varepsilon_{\Sigma fr\, cond}$ and formal maximal strains $\varepsilon_{\Sigma max}$ for the two- phase two-modal (in the left column) and single-modal (in the right column) PC aggregates α-Fe with small- and large-angle GB with allowance for Table 8 at $T$=300K.

The results produced in this section – deformation curves in Fig. 10, 11 on a basis of the strength and plastic characteristics $(\sigma_{(\Sigma)y}, \sigma_{(\Sigma)S}, \varepsilon_{\Sigma fr\, cond}, \varepsilon_{\Sigma max})$ calculated in Tables 8, 9 – confirm the evaluation of an influence both of GB regions and multi-modality (presented in the Sections 8 and 9, see Table 5) on the behavior of these quantities. In particular, in the case of small-angled GBs (corresponding to the small porous part) the behavior of the deformation curves is changed insignificantly for single-modal two-phase PC aggregates with decreasing of the strength factors. An addition of composite second-mode grains to NC aggregates (initially with the first phase grains for $d \leq 150\, nm$) reduces both the plasticity and strength, as well as increases the strength $(\sigma_{\Sigma y}, \sigma_{\Sigma S})$ starting from SMC and UFG samples till CG PC aggregates. For the large-angle GB the changes are more radical with preservation of the same tendencies. An increase in plasticity by 1.5-2.5 times appears to be essential for SMC and UFG two-mode PC samples, as compared to single-mode ones. We observe that the presence of a second-phase makes stress-strain curves very similar to those obtained experimentally in [49] for SMC and UFG Armco-Fe samples, thus providing the correctness of the plots in Figs. 10, 11 at the NC regions ($d \leq 100\, nm$) for PC α-Fe samples. The latter still awaits an experimental support.

We stress, first of all, that an adaptation of the theoretical model in order to account for different dislocation ensembles under quasi-static deformation with corresponding probabilities of its emergence with passing from equidistant spectra of the crystallite energy to more general spectra permits one to essentially specify the form of the deformation curves $\sigma_\Sigma - \varepsilon$. Second, the more realistic case for the deformation curves for two-phase PC samples implies the different changes of the values $\varepsilon_1, \varepsilon_2$ and $\varepsilon_P$ due to non-homogeneous character of deformations for different phases and modes of grains within the general integral FS and strains (47).

### 13. Summary and discussion

A statistical approach used to derive the generalized Hall–Petch relation for yield strength, including an analytic form of stress-strain dependence has been developed on a basis of analyzing the mechanical energy spectrum of each crystallite in a single-mode (of diameter $d$) polycrystalline aggregate under qua-si-static loading with a constant rate $\dot{\varepsilon}$. For a fixed value of PD, a crystallite spectrum consisting of discrete energy levels $E_d^0(\varepsilon), E_d^1(\varepsilon), E_d^2(\varepsilon), \ldots, E_d^n(\varepsilon), \ldots, E_d^N(\varepsilon)$ is considered in the equidistant approximation with a step equal to the energy of a unit dislocation (in the most probable dislocation ensembles for a given phase in a studied material) and corresponding to the emergence of a dislocation with $E_d^n(\varepsilon)$, for a Burgers vector of minimal length, and with (*n*+1) atoms (*n* elementary segments) in the axis. A scenario is proposed for implementing a rectilinear edge dislocation of length $L = Nb$ in a crystallite with a cubic CL under constant tension through the formation within a thermal-fluctuation mechanism of a sequence of 0D-defects – nanopores, being by the zone of localized plasticity for a time interval, $t_d \approx N \cdot b / v_s$ much lesser ($t_d \ll \Delta t_0$) than $\Delta t_0 = b/(\dot{\varepsilon} d)$ in (7) between the act of plastic deformation in the crystallite. Thus, we may select two time scaling: the fast one $t_d$ for forming of a dislocation and the slow one $\Delta t_0$ for enumerating the PD acts. The process of quasi-static PD of the crystallite and PC aggregate represents the sequence of equilibrium processes being changed at changing of ε by skipping from one to another. In a state of thermodynamic quasi-equilibrium, the probabilities of finding a crystallite in a state characterizing by the vector $(m_1, m_2, \ldots, m_n, \ldots m_N)$ having $m_1$ unit dislocations (including nanopores), $m_2$ dislocations with 3 atoms on the axes, ..., $m_n$ dislocations with (*n*+*1*) atoms on the axes, and then up to $m_N$ maximal rectilinear dislocations passing through the center of the crystallographic equatorial plane, are given by the Boltzmann distribution (9), (10), with the scale energy factor $M(\varepsilon)$ relative to the energy $E_d^N(\varepsilon)$ of the maximal dislocation. The estimation (7) of the minimal time interval for the emergence of a dislocation in a crystallite under PD allows one to estimate the number of dislocations (8) with an accumulated PD, *ε*, whereas the average number of atoms (13) (segments of the average dislocation) on a dislocation axis, obtained from the Bose–Einstein distribution, leads to the analytic representation (17) for equilibrium scalar dislocation density, which coincides with the experimental numerical values (18) within the limits of CG and NC materials. An assumption of Taylor's strain hardening mechanism validity (19) in all grain size regions leads to the representation (20) for flow stress $\sigma(\varepsilon)$, starting from the first (crystalline) phase of a polycrystalline aggregate without a special texture. From the generalized flow stress and Hall–Petch laws for yield strength $\sigma_y$, exact expressions are obtained for the maximum flow stress (23) and extreme grain size $d_0(\varepsilon, T)$ (22), within the nanometer range. The value of $d_0(\varepsilon, T)$ grows with an increase in PDs and decrease in temperatures. In the limit of CG aggregates, the well-known normal Hall–Petch law follows from (20), which allows one to refine the polyhedral parameter $m_0$ (24) by using experimental data (see [38, 39] for details). The graphical representations (Fig. 5) of the established HP law for one-phase PC aggregates with BCC (α- Fe), FCC (Cu, Al, Ni) and HCP (α-Ti, Zr) crystal lattices with closely-packed grains at T=300K show a very close coincidence between the theoretical and experimental data both for the extremal grains $d_0(0.002, 300)$ and for the maximum $\sigma_m(0.002)$ for these materials.

A quasi-particle interpretation of the quantization of the energy of crystallite in a PC aggregate under PD is given. In this interpretation the quantum of energy of unit dislocation, tentatively referred to as a *dislocon,* appears to be a composite (short-lived) particle consisting of acoustic phonons near two atoms (at the instant of destruction for the bond between the atoms, followed by a creation of a nanopore), which exits the nodes of the crystal lattice under PD. This idea provides an analytic approach to the description and evolution of the Chernov– Luders PD macroband, recently initiated by [50], in connection with the observed acoustic emission [51, 52]. The quasi-particle interpretation adds a significant argument in flavor of the concept for the origin of 1*D*-defects (dislocations) in terms of 0*D*-defects (nanopores, vacancies) under the PD process. This interpretation has enabled us to justify the distribution (10) of energy states in crystallites $\{P(E_n, \varepsilon)\}$ by considering an assembly of dislocons as a dislocon gas, which is shown to have a Boltzmann-type distribution for its number (concentration) in (31) between the PD acts, once a discrete change of crystallite energy under PD is ignored.

The study of the temperature behavior of the strength characteristics reveals a new effect*,* which we called a *temperature-dimension effect* initially established in the one-phase model of a single-mode PC material. It means that with a growth of temperature the extremal grain size $d_0(\varepsilon, T)$ decreases, and the yield strength $\sigma_y$, including its maximum $\sigma_m$ increases for all the grains with $d < d_1$ (with a new criti-cal size of grains $d_1$ (35): $d_1 \approx 3d_0$), whereas $\sigma_y$ decreases for subcritical grains with $d > d_1$ as usual in the UFG, FG and CG limits and it has been demonstrated theoretically for pure Al in Fig. 6.

For the stress-strain curves in one-phase single-mode α-Fe shown in Fig.7 (for $d=10^{-3}$; $10^{-4}$; $10^{-6}$m; 150 nm; $d=d_0=23.6$ nm;10nm for $T=$ 300 K), the formal stress maximum $\sigma_{max}$ increases with a decrease in the linear grain size, shifting to the region of smaller strains $\varepsilon_m$, starting from $\varepsilon_m=0.5$ in CG materials and reaches its absolute maximum in the NC region, with $(\varepsilon_m, d_m) = (0.2, 40.6)$nm), $\sigma_{max}(\varepsilon_m, d_m, 300) = 13.27$ GPa at $T=300$ K. Continuing a decrease for the grain size $d<40.6$ nm, the maximum $\sigma_{max}$ decreases significantly, along with the plasticity region. The validity of the Backofen–Considére condition (39) of fracture (established using a strain hardening coefficient $\theta(\varepsilon)$) is implemented for the conditional fracture strain $\varepsilon_{fr.cond.}(d)$ in (40), thereby making the maxima $\sigma_{max}(d_m)$ physically unreachable due to $\varepsilon_{fr\,cond}(d_m) \approx 0.135 < \varepsilon_m$.

In order to account for the GB influence to integral strength and plastic characteristics, the one-phase model of PC aggregates has been augmented by a two-phase model with composite grains (Fig. 3), which consists of the first (solid or pure crystalline) phase (in the terminology of [5]) enveloped by a second (weak or GB) phase. The latter includes a crystalline (fragmentary) part of smaller size values and some pores necessarily present between the grains. The analytic presence and input of the second phase into the integral stress $\sigma_\Sigma(\varepsilon)$ (42), (45) has been taken into account additively, with a positive input from the fragmentary and negative parts from the porous part of the second phase, with respective weights $(1 - \kappa_1), (\kappa_1 - \kappa_2), \kappa_2$, calculated according to the average size of first-phase and second-phase grains, as well as of pores, by using the rule (41), (43), (44). The relations (42) and (45) (see also Footnote 10) with the porous structure alone at the second phase, as well as its natural generalization for the integral $\sigma_{\Sigma dis} - \varepsilon$ dependence (47) to the case of three-phase model with third-phase composite grains playing the role of dispersion hardening particles (from other compounds, Cu, in [41]) are the principal theoretical results of the two-phase model concept in PC. It is shown using an example of α-Fe at $T=300$ K that the stress-strain curves, in particular, the value of maximum, strongly depend on the GB part of an aggregate, and the formal maximal value $\sigma_{\Sigma max}(d_m, 300) = 4.18$ GPa decreases by 3.2 times with respect to $\sigma_{max}(d_m,300) =13.27$ GPa with the respective weights $(1 - \kappa_1) = 0.577$, $\kappa_1 = 0.423$, whereas the actual maximum (in the 2-phase model) $\bar{\sigma}_{\Sigma max}(d_m,300)$ is already reached at $\bar{\varepsilon}_m > \varepsilon_m$, which, however, cannot be implemented because of the Backofen–Considére condition (49) with $\varepsilon_{\Sigma fr\,cond}(d_m, 300) > \varepsilon_{fr\,cond}(d_m, 300) \approx 0.135$. The reachable absolute maximum (in $d, \varepsilon$) (the conditional ultimate stress in question) calculated at $\varepsilon_{fr\,cond} \approx 0.135$ is evaluated as $\sigma_S = 3.37$ GPa.

It is shown that the introduction of the second-phase part into the single-mode PC aggregate model influences both the form of Hall–Petch law for $\sigma_{\Sigma y} = f(d^{-1/2})$, at the value of the extremal average grain size $d_{\Sigma 0}$ and the maximum of yield strength $\sigma_{\Sigma m}$, as compared to the one-phase model of PC aggregates with BCC (α- Fe), FCC (Cu, Al, Ni) and HCP (α-Ti, Zr). Namely, from Table 6 and Figures 8 it follows that for a small-angle GB (i.e., for $d_{Ps} = 0.02d$ and its weight 0.038) that $d_{\Sigma 0}$ and $\sigma_{\Sigma m}$ become smaller than $d_0$ and $\sigma_m$ respectively. For a large-angle GB (i.e., for $d_{Pg} = 0.1d$ and its weight 0.167) the same tendencies are realized for all materials with a decrease in $\sigma_{\Sigma m}$, approximately by 20%. In the case of constant pores, $d_P = d_0$, the extremal grain size values are shifted into the SMC region with a multiple decrease in the $\sigma_{\Sigma m}$ and the lowest boundary $d_{LB}$ arises for the existence of two-phases single-mode PC samples. These results are complete agreement with the actual behavior of experimental two-phase PC aggregates.

The study of the second-phase part, introducing to the temperature behavior of two-phase PC aggregates, using the example of Al, presented by Table 7 and Figure 9, in comparison with the same study for the one-phase model reveals a conservation of TDE for both small- and large-angle GBs with a decrease in the extremal size values $d_{\Sigma 0}$, $d_{\Sigma 1}$ and $\sigma_{\Sigma m}$ as compared, respectively, with $d_0$, $d_1$ and $\sigma_m$ within the temperature range [150K, 350K]. In the case of constant pores (Fig. 9c) with $d_P = d_0(300)$ this effect is completely neutralized and the PC material obeys the standard $T$-behavior from CG down to SMC aggregates. The reason for the above neutralization to appear is due to shifting from the extremal NC grain region for a composite model of grains to the subcritical SMC region, so that the new extremal values $d_{\Sigma 0}$ already belong to the range of 125–175 nm. The predicted TDE in two-phase single- and multi-mode PC aggregates is in need (!) of experimental verification. This verification is more readily implemented start-ing from low temperatures, however, in the framework of the same phase (not to be confused with the above solid or weak phase).

Finally, the influence of multimodality and GB values on the stress-strain curves $\sigma_\Sigma = \sigma_\Sigma(\varepsilon, d_1, d_2, T)$ and $\sigma_\Sigma = \sigma_\Sigma(\varepsilon, d_1, 0, T)$ calculated according to the basic formula (47) has been studied by applying to α-Fe two-phase PC aggregates for CG, $d=10^{-3}$ m, $10^{-4}$ m, UFG, $d=1$ μm, SMC, $d=150$nm, NC, $d=d_0=23.6$ nm and 10 nm, for the second mode grains with $d_2 = 0.5d$ and the weight $f_2 = 30\%$. The results presented on a basis of the strength and plastic characteristics ($\sigma_{(\Sigma)y}, \sigma_{(\Sigma)S}, \varepsilon_{\Sigma fr\ cond}, \varepsilon_{\Sigma max}$) in Tables 8, 9 and Figs. 10, 11 for small- and large-angle GBs with the weights of the phases given by Eqs. (50), (51), confirm the evaluation of an influence of GB regions and multi-modality on the behavior of these quantities. In particular, in the case of small-angle GBs the behavior of the deformation curves is changes insignificantly for single-mode two-phase PC aggregates with a decrease in the strength factors. An addition of composite second-mode grains into the NC aggregates (initially, with the first phase grains for $d \leq 150$ nm) reduces plasticity and strength, as well as increases strength ($\sigma_{\Sigma y}, \sigma_{\Sigma S}$), from SMC and UFG samples down to CG PC aggregates. For large-angle GBs, the change is more significant but has the same tendencies. An increase in in plasticity by 1.5-2.5 times appear to be essential for SMC and UFG two-mode PC samples, as compared to single-mode ones.

Discussing the results of our research, we stress, in reference with the obtained analytic relations, which establish an immediate connection among the microstructure parameters of equilibrium PC materials: size of crystallite modes, grain boundary magnitude (size and weight of nano- and micro-pores), Burgers vector, weight parts of the phases, texture with its integral strengthened and plastic characteristics under quasi-static deformation, that they in principle cannot be obtained, it would seem, without an application of statistical analysis developed in the research. The presence of experimentally well-established extremal values of yield strength $\sigma_y$, ultimate stress $\sigma_S$ to be reached in the respective critical average diameters of grains $d_0, d_m$ and strain $\varepsilon_{fr\ cond}$, respectively, in the Hall–Petch relations and the deformation curves with allowance for the fact that the physical dimensions of $\sigma_y$, $\sigma_S$ correspond to the dimension of energy density, $[\sigma_y]=[\sigma_S]=1$eV·m$^{-3}$, gives the unambiguous conclusion that the energy of a grain, being a basic component of a PC material, must change in a discrete manner under PDs, i.e., be oscillators. It is naturally based on the discrete (atomic) structure of the grains and translational (locally) invariance of its CL, and is accompanied by the emergence of defects, including dislocations, being most stable in metallic PC aggregates. As the consequence an emergence of a defect at a given place of CL requires the localization in this place of the minimal energy quantum to generate a defect under external loading. This quantum almost coincides with the (experimental) activation energy of an atom at the process of diffusion for a given PC material. It has been established that this portion corresponds to the energy of a unit dislocation from the ensemble of most probable dislocations. This discrete energy portion, as one follows the idea of wave-particle duality, becomes very useful and important in the form of a quasi-particle, dislocon. The dislocon seems to be a natural carrier of interaction both between the deformed and (un)deformed parts of a crystallite and between neighboring crystallites, being periodically absorbed by the grain's CL, as well as by the GB or is decayed on phonon compounds, thereby locally changing the temperature in the sample. Such quasi-particles seem to be adequate candidates for implementing energy fluxes in PC materials under PDs, especially in describing such macroscopic effects of localized deformations as the Chernov– Lüders macroband and the Portevin–Le Chatelier effect [50, 53].

Among the peculiarities of the model and the problems that require further research, we can point out the following:
1. Necessity of a more precise analytic determination of the value of the *polyhedral parameter* $m_0$, so far determined only in the CG limit (24);
2. Allowance for other dislocation ensembles, in addition to the classically most probable ensemble under PDs, for instance, by the construction of a Hamiltonian which leads to a non-equidistant energy spectrum, and of an analog of the evolution equation (a kind of the Schrödinger equation) whose solution could be determine the probability of defects to originate – dislocations – enlarging the Bose–Einstein distribution (13) in order to account for them in the integral strain-stress law $\sigma_\Sigma - \varepsilon$, possibly outside the equilibrium state;

3. Allowance for the texture of a grain distribution in a PC aggregate by extending the isotropic distribution (the absence of a texture) to the case of special distributions, arising at SPD by ECAP, torsion under pressure, or magnetron sputtering (used for thin coverings);
4. Research for the influence of the twinning process, within a disclination-dislocation deformation mechanism of twin formation relative to the integral curve $\sigma_\Sigma - \varepsilon$;
5. Derivation of the integral laws $\sigma_\Sigma - \varepsilon$ for multi-phase PC aggregates on a basis of Fe and doping elements C, Cr, V, Mn, etc., simulating the behavior, e.g., ferritic-pearlitic steels (considered, e.g., in [54]), while obtaining 3D bulk samples with a lengthy yield surface, large yield strength and ultimate stress due to multi-phase (in the usual sense) multi-modality, values of GB and input from dispersion hardening;
6. Simulation of composite PC aggregates, in which the second-phase with a sufficiently large-angle GB plays the role of a matrix, whereas the role of a filling material (first-phase grains) is played by high-strength fibers, high-plastic compounds[13], or high-melting particles of various dispersity, not dissolving in the basic metal (such as high-melting oxides, nitrides, borides, carbides), with a modification of the strength and plastic characteristics, as well as the high-temperature strength, of the final PC aggregate.

We emphasize, once again, that the TDE effect is in need of experimental verification, which has to be the criterion of validity of the suggested theory. The theoretical model implies evident perspectives of its application to new PC composite materials, including those obtained by additive technologies, in the aircraft and cosmic industry, and has been tested experimentally using samples of the BT1-0 α-Ti alloy and UFG PC samples [48]**.**

The author is grateful to E.V. Shilko, Yu.P. Sharkeev, I.A. Ditenberg for discussions of various aspects of the present research. He also thanks Academician V.E. Panin for useful criticism, as well as the chairman and participants of physical seminars at the Institute of Strength Physics and Materials Science SB RAS, where the idea of this work came to life, and its development was stimulated in the course of fruitful discussions.

The work has been supported under the Program of fundamental research at the State Academy of Sciences for 2013-2020.

---

[13] This has been recently implemented, e.g., in composite Silumin of weight ≤ 40%, with a filling of Sn [55], which significantly increases the plasticity region as compared to pure Silumin with no loss of strength properties.